\documentclass[journal]{IEEEtran}
\pdfoutput=1
\usepackage{cite}
\usepackage{amsmath,amssymb,amsfonts}
\usepackage{mathtools}
\usepackage{amsthm}
\usepackage{bm}
\usepackage{graphicx}
\usepackage{array}
\usepackage[caption=false,font=footnotesize]{subfig}
\usepackage{algorithm}
\usepackage{algorithmic}
\usepackage{booktabs}
\usepackage{multirow}
\usepackage{makecell}
\usepackage{xcolor}
\DeclareRobustCommand{\introchange}[1]{{#1}}
\DeclareRobustCommand{\modelchange}[1]{{#1}}
\DeclareRobustCommand{\structurechange}[1]{{#1}}
\DeclareRobustCommand{\methodchange}[1]{{#1}}
\makeatletter
\let\TSPoriginalfloatmakebox\float@makebox
\renewcommand{\float@makebox}[1]{\color@vbox\normalcolor\TSPoriginalfloatmakebox{#1}\color@endbox}
\makeatother
\usepackage{url}
\usepackage{flushend}
\usepackage{stfloats}
\graphicspath{{figs/}}

\allowdisplaybreaks[1]

\makeatletter
\providecommand{\captionof}[2]{\begingroup\def\@captype{#1}\caption{#2}\endgroup}
\makeatother

\newtheoremstyle{paperthm}
{6pt plus 2pt minus 2pt}
{6pt plus 2pt minus 2pt}
{\normalfont}
{}
{\bfseries}
{.}
{0.5em}
{\thmname{#1}\thmnumber{ #2}\thmnote{ \emph{(#3)}}}
\theoremstyle{paperthm}
\newtheorem{theorem}{Theorem}
\newtheorem{lemma}{Lemma}
\newtheorem{proposition}{Proposition}
\newtheorem{corollary}{Corollary}

\newtheorem{assumption}{Assumption}
\newtheorem{remark}{Remark}

\newcommand{\C}{\mathbb{C}}
\newcommand{\E}{\mathbb{E}}
\newcommand{\rank}{\operatorname{rank}}
\newcommand{\spanop}{\operatorname{span}}
\newcommand{\diag}{\operatorname{diag}}
\newcommand{\tr}{\operatorname{tr}}
\newcommand{\range}{\operatorname{range}}

\newcommand{\RevValidationQRange}{5\text{--}6}
\newcommand{\RevValidationGapMax}{0.440}

\newcommand{\RevScalingQMax}{7}
\newcommand{\RevScalingSmallTime}{0.09}

\newcommand{\RevScalingLargeTime}{72.50}
\newcommand{\RevScalingLargeFraction}{0.53}
\newcommand{\RevDriftSmall}{2.71}

\newcommand{\RevHeldCertSmall}{0.66}
\newcommand{\RevDriftMiddle}{0.08}

\newcommand{\RevHeldCertMiddle}{0.50}
\newcommand{\RevDriftLarge}{0.01}

\newcommand{\RevHeldCertLarge}{0.49}

\newcommand{\RevPilotLargeFullTime}{202.56}
\newcommand{\RevPilotLargeDynamicTime}{9.549}
\newcommand{\RevPilotLargeFullRatio}{21.4}
\newcommand{\RevPilotLargeBMTime}{57.17}
\newcommand{\RevPilotLargeBMRatio}{6.9}

\newcommand{\RevPilotScenes}{8}

\newcommand{\StrongMThreeDynamicQ}{9}
\newcommand{\StrongMThreeTailQ}{35\text{--}39}
\newcommand{\StrongMThreeRatio}{5.4}
\newcommand{\StrongMFourDynamicQ}{13}
\newcommand{\StrongMFourTailQ}{41\text{--}42}
\newcommand{\StrongMFourRatio}{4.2}
\newcommand{\StrongPublicCount}{64}
\newcommand{\StrongPublicScenes}{8}
\newcommand{\StrongPublicSuccess}{51}
\newcommand{\StrongPublicDynamicSuccess}{64}
\newcommand{\StrongPublicRatioMin}{11.5}
\newcommand{\StrongPublicRatioMax}{26.3}

\newcommand{\StrongPublicBundleSize}{16}
\newcommand{\StrongPublicPastSize}{8}

\newcommand{\StrongMechanismCount}{70}
\newcommand{\StrongTailSuccess}{60}
\newcommand{\StrongDynamicExpanded}{68}
\newcommand{\StrongStrictSteps}{258}

\newcommand{\StrongStrictMinSeparation}{0.019}

\newcommand{\StrongPublicHardSmallGap}{2.27}

\newcommand{\StrongPublicHardLargeGap}{13.45}

\newcommand{\ReceiveTightMin}{2.0}
\newcommand{\ReceiveTightMax}{3.3}

\newcommand{\ReceiveMainReferenceCertMax}{0.551}
\newcommand{\ReceiveObsSmallExcess}{7.9}
\newcommand{\ReceiveObsLargeExcess}{29.1}
\newcommand{\ReceiveObsDynamicMax}{0.36}

\newcommand{\ReceiveObsCount}{114}

\newcommand{\ReceiveObsTwoOBSQ}{8\text{--}10}
\newcommand{\ReceiveObsTwoFisherQ}{4\text{--}8}

\newcommand{\ReceiveRandomSmallRatio}{14.6}

\newcommand{\ReceiveRandomMiddleRatio}{18.5}

\newcommand{\ReceiveRandomLargeRatio}{20.7}

\newcommand{\ReceiveRandomMultiRatio}{4.8}

\newcommand{\ReceiveThreadsLargeOneRatio}{25.4}
\newcommand{\ReceiveThreadsLargeFourRatio}{14.9}
\newcommand{\ReceiveThreadsLargeEightRatio}{13.2}

\newcommand{\ReceiveMismatchMax}{6.66}
\newcommand{\MultiBudgetDynamicSixGap}{0.40}
\newcommand{\MultiBudgetDpssSixGap}{5.65}

\newcommand{\SampLHSMax}{1.36}
\newcommand{\SampLHSMedianMin}{0.12}
\newcommand{\SampLHSMedianMax}{1.04}
\newcommand{\SampLHSAllMax}{1.67}

\newcommand{\SampMCMax}{81.0}
\newcommand{\SampStabBCRB}{0.34}

\begin{document}

\title{Certified Dynamic Subspace Optimization \\for Bayesian ISAC Design}

\author{Zongyao Zhao,~\IEEEmembership{Member,~IEEE}, Cheng-Jie Zhao,~\IEEEmembership{Graduate Student Member,~IEEE}, Zhaolin Wang,~\IEEEmembership{Member,~IEEE},
 \\Qingjiang Shi,~\IEEEmembership{Senior Member,~IEEE}, and Yuanwei Liu,~\IEEEmembership{Fellow,~IEEE}

\thanks{Corresponding author: Yuanwei Liu.}

\thanks{Z. Zhao, C.-J. Zhao, Z. Wang, and Y. Liu are with the Department of Electrical and Computer Engineering, The University of Hong Kong, Hong Kong S.A.R., China. (E-mails: zongyao@hku.hk, chengjie\_zhao@connect.hku.hk, zhaolin.wang@hku.hk, yuanwei@hku.hk)}

\thanks{Q. Shi is with the School of Computer Science and Technology, Tongji University, Shanghai, China. (E-mail: shiqj@tongji.edu.cn)}
}

\maketitle

\begin{abstract}
Full-space covariance optimization is costly in large-array Bayesian
integrated sensing and communication (ISAC). A fixed transmit subspace can
reduce this cost, but the resulting optimality loss depends on the posterior
distribution. In this paper, a certified dynamic
subspace method is developed for angle Bayesian Cram\'{e}r--Rao bound (BCRB)
minimization under joint angle and complex-gain uncertainty. For uniform
linear arrays, exact response preservation over a nondegenerate posterior
angular interval requires the full antenna space. Subspace sufficiency is
therefore assessed through full-space objective-gap bounds. A
communication-feasible basis is initialized from the user-channel span and
local BCRB sensitivity. Restricted semidefinite programs are then solved.
Primal restoration and dual repair yield gap bounds that remain valid after
lossless user-beam recovery. When an accurate restricted solve does not
certify the requested tolerance, the basis is expanded using a direction
identified from the covariance dual slacks. Under a strict communication-power
margin, joint uniqueness of the restricted optimal dual multipliers is
established. With exact restricted solves, this uniqueness guarantees strict
BCRB improvement after certificate-driven expansion. Numerical results
validate the bounds against full-space references for the sampled problems.
Compared with a fixed Fisher ordering, the proposed method can reach the
same certificate target with fewer dimensions. On
large arrays, total runtime is reduced relative to the full-space and
factorized baselines at matched certificate targets.
\end{abstract}


\section{Introduction}

\IEEEPARstart{I}{ntegrated} sensing and communication (ISAC) is expected to be
an enabling technology for future wireless systems~\cite{Liu2020RoadAhead,Zhang2021Overview,Liu2022JSAC}.
Communication infrastructure, spectrum, and waveforms are reused for both data
transmission and sensing. Communication quality of service must be protected
while useful sensing information is provided. Large antenna arrays improve
spatial resolution and array gain, particularly at millimeter-wave and
terahertz frequencies. However, transmit design becomes increasingly
expensive as the number of antennas grows.

Representative ISAC transmit designs employ diverse sensing criteria.
Response-oriented formulations optimize beampattern and radar-gain metrics~\cite{Hassanien2016DFRC,Liu2018,Liu2020JointBeamforming,Hua2023}.
Detection-probability criteria are also used in multi-target and backscatter
ISAC designs~\cite{Zhao2025MultiTargetDetection,Zhao2024Backscatter}.
Estimation-oriented formulations instead use criteria such as the
\introchange{Cram\'{e}r--Rao bound (CRB)}, Fisher information matrix (FIM), or mean-squared error~\cite{Liu2022CRB,Zhao2024CRB,Wang2024ExtendedCRB,Wu2024GlobalISAC,Lu2024RandomISAC}.
Target uncertainty is addressed through prior-aware or Bayesian formulations~\cite{Zhao2025BayesianBeamforming,AttiahYu2024BCRB,XuZhang2024PriorISAC}.

In Bayesian ISAC design, sensing targets provide no dedicated channel state
information (CSI) feedback. Their angles and reflection coefficients are
inferred from tracking filters, maps, or previous snapshots and are therefore
described by posterior distributions~\cite{Zhao2025BayesianBeamforming,AttiahYu2024BCRB}.
User channels, in contrast, are acquired before downlink transmission through
pilot training and CSI feedback. Accordingly, communication channels are treated
as known, whereas sensing parameters are treated statistically.
\introchange{For angle estimation, the CRB is used as a design criterion. Under target uncertainty, its Bayesian counterpart (BCRB)
incorporates both observation and prior information~\cite{VanTrees2001,VanTrees2002Array}.
For large arrays, high-dimensional matrix variables make transmit covariance
optimization expensive. This cost can be reduced by restricting the variables
to a low-dimensional transmit subspace.}
The central question is whether the resulting design attains an objective
value sufficiently close to the full-space optimum. This must be certified
without solving the full-space Bayesian covariance problem.

Exact subspace reductions are available when the relevant response atoms are
known. With fixed user channels and target responses, the communication and
CRB terms depend on the transmit covariance through channel, steering, and
derivative-steering responses. Projection onto their joint span preserves
these terms without increasing transmit power~\cite{Zhao2026NearFieldET}.
\introchange{Related low-dimensional structures have been obtained from stationary-point,
Karush--Kuhn--Tucker (KKT), and duality analyses.
These include the low-dimensional subspace (LDS) and optimal beamforming
structure (OBS) results of~\cite{ZhaoShi2025LDS,Zhao2026ISACLDS,Fang2025ISACOBS}.}
Beamspace, hybrid precoding, joint spatial division and multiplexing (JSDM),
and low-complexity solver structures provide complementary large-array
reductions~\cite{ElAyach2014Sparse,Alkhateeb2014Hybrid,Adhikary2013JSDM,Wang2025ISACSubspace,Zhao2023RWMMSE}.

\introchange{For Bayesian ISAC, posterior sensing directions are characterized
in~\cite{AttiahYu2024BCRB,AttiahYu2025Duality}.
An uplink--downlink duality with potentially indefinite virtual noise is
established in~\cite{AttiahYu2025Duality}.
With $K$ users, the required number of beams is also bounded for sensing metrics
expressed through $d$ real quadratic forms~\cite{AttiahYu2026HowMany}.}

\introchange{For uniform linear arrays (ULAs), Section~\ref{sec:theory} shows that exact
response preservation over a nondegenerate posterior spatial-frequency
interval requires the full antenna space. Under strict feasibility and information regularity,
an optimal design is attained on a support whose dimension is bounded by
the user and target counts~\cite{AttiahYu2026HowMany}.
The bound is independent of the antenna count, but the support directions
must be determined for each posterior instance. A candidate subspace is therefore assessed by bounding its objective gap
to the full-space semidefinite program (SDP).}
\introchange{Table~\ref{tab:positioning} compares these subspace structures and support
existence bounds with the objective-gap certificate developed here.}

\begin{table}[!t]
    \centering
    \caption{Subspace structures and guarantees.}
    \label{tab:positioning}
    {\footnotesize
    \setlength{\tabcolsep}{1.5pt}
    \renewcommand{\arraystretch}{1.1}
    \begin{tabular*}{\columnwidth}{@{\extracolsep{\fill}}>{\leavevmode}p{0.21\columnwidth}>{\leavevmode}p{0.13\columnwidth}>{\leavevmode}p{0.42\columnwidth}>{\leavevmode}p{0.16\columnwidth}@{}}
    \toprule
    Method & \introchange{Sensing} & Reduced structure & Guarantee\\
    \midrule
    \mbox{JSDM~\cite{Adhikary2013JSDM}} & none & long-term covariance & \introchange{structure}\\
    \mbox{LDS~\cite{ZhaoShi2025LDS}} & none & channel span & \introchange{structure}\\
    \mbox{ISAC-LDS~\cite{Zhao2026ISACLDS}} & \introchange{extended} &
    \makecell[l]{channel-range user blocks and\\null-space sensing block} & \introchange{structure}\\
    \mbox{OBS~\cite{Fang2025ISACOBS}} & \introchange{nominal} & $\spanop[\bm H,\bm A,\dot{\bm A}]$ & \introchange{structure}\\
    \introchange{\mbox{Beams~\cite{AttiahYu2026HowMany}}} & \introchange{posterior} & \introchange{$\le\lfloor\sqrt{K^2+d}\rfloor$ beams} & \introchange{existence}\\
    \textbf{This work} & posterior & \introchange{certified adaptive transmit span} &
    \introchange{gap bound}\\
    \bottomrule
    \end{tabular*}
    \par}
\end{table}

A static posterior basis can be constructed from directions with large
average response energy. A direction with moderate energy may nevertheless be
important for a weakly informed angle.
BCRB sensitivity depends on the transmit covariance through the Bayesian
information matrix (BIM).
A basis selected at a reference covariance must therefore be checked after
reduced optimization redistributes power.

\begin{figure*}[!t]
 \centering
 \includegraphics[width=\textwidth]{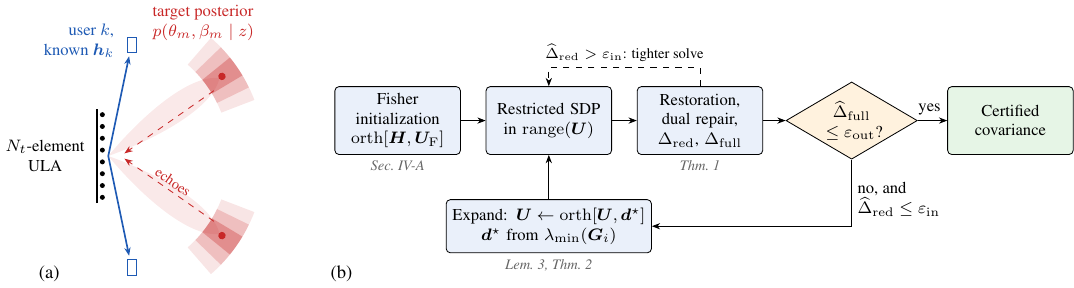}
 \caption{System and method overview. (a) Downlink ISAC with known user
 channels and posterior target uncertainty. (b) Certified dynamic subspace
 optimization.}
 \label{fig:system_overview}
\end{figure*}

\introchange{To address this gap, a shared transmit basis is initialized from the channel
span and local BCRB sensitivity. Restricted SDPs retain the joint-posterior
information map and original signal-to-interference-plus-noise ratio (SINR)
constraints. After accurate reduced solves, spectral generation is used to
expand the basis when the full-space certificate exceeds tolerance~\cite{HelmbergRendl2000,OskoorouchiGoffin2007,LiaoDingZheng2026,AngellMcCallum2024}.
Here, dynamic refers to basis expansion within a fixed posterior instance.
Under a strict communication-power margin, joint dual uniqueness is
established from the SINR block structure. After an exact restricted solve,
the selected direction makes the unique optimal dual solution infeasible for
the enlarged dual problem. A strictly lower BCRB is therefore attained after
an exact solve of the enlarged primal problem.
The returned feasible covariance carries a full-space
objective-gap bound (Fig.~\ref{fig:system_overview}(b)).}

The main contributions are summarized as follows.
\begin{IEEEitemize}
    \widowpenalty=10000
    \item \introchange{A dynamic transmit subspace method is developed for Bayesian
    ISAC covariance optimization. A communication-feasible basis is initialized
    from the user-channel span and local BCRB sensitivity. The basis is expanded
    as needed using covariance dual slacks.}

    \item \introchange{Computable reduced- and full-space objective-gap bounds are derived
    from repaired dual variables. Analytic optimization of the shared power
    multiplier tightens these bounds. The full-space certificate remains valid
    for restored feasible covariances and after lossless user-beam recovery.}

    \item \introchange{Under a strict communication-power margin, joint uniqueness
    of the optimal Schur, SINR, and power multipliers is established.
    With exact restricted solves, certificate-driven expansion is shown to
    strictly improve the restricted optimal BCRB while preserving feasibility.}

    \item \introchange{At matched certificate targets, dynamic expansion can use fewer
    dimensions than fixed Fisher ordering requires in multi-target experiments.
    For large-array sampled SDPs, total runtimes are also lower than those of
    full-space MOSEK and the factorized baseline.}
\end{IEEEitemize}

\emph{Notation\introchange{.}} Bold lowercase and uppercase letters denote vectors and
matrices. $(\cdot)^H$, $\tr(\cdot)$, $\rank(\cdot)$, and $\range(\cdot)$ denote
conjugate transpose, trace, rank, and column space, respectively.
$\spanop(\cdot)$ denotes linear span.
$\operatorname{orth}(\cdot)$ returns an orthonormal column-space basis.
$\diag(\cdot)$ and $\operatorname{blkdiag}(\cdot)$ construct diagonal and
block-diagonal matrices.
$\|\cdot\|_2$ is the Euclidean or spectral norm, and $\|\cdot\|_F$ is the
Frobenius norm. Eigenvalues are in nonincreasing order, $\E[\cdot\mid z]$ is
conditional expectation, and $\mathcal{CN}(\bm\mu,\bm R)$ denotes a proper
complex Gaussian distribution.

Section~\ref{sec:system} introduces the system model and problem formulation,
and
Section~\ref{sec:theory} \introchange{examines exact response preservation and
instance-specific optimal support.} Section~\ref{sec:reduced_isac_optimization} develops the certified
dynamic method and analyzes its complexity. Numerical results are reported in
Section~\ref{sec:experiments}, followed by discussion and conclusions in
Sections~\ref{sec:discussion} and~\ref{sec:conclusion}, respectively.
\introchange{Additional proofs and implementation details appear in the supplement.}

\section{Bayesian ISAC Model and Problem Formulation}\label{sec:system}

\subsection{Signal and Posterior State Model}

We consider a narrowband downlink ISAC transmitter equipped with an
$N_t$-element half-wavelength-spaced ULA. The transmitter serves $K$
single-antenna users and estimates the angles of $M$ targets, as illustrated
in Fig.~\ref{fig:system_overview}(a). The transmitted
signal at sample $n$ is
\begin{equation}\label{eq:tx_signal}
    \bm x[n]=\bm W\bm s[n],
\end{equation}
where $\bm s[n]\in\C^L$, $L\ge K$, satisfies
$\E[\bm s[n]\bm s^H[n]]=\bm I_L$, and
$\bm W=[\bm w_1,\ldots,\bm w_K,\bm W_s]\in\C^{N_t\times L}$. The first $K$
columns carry user data and $\bm W_s$ contains the probing precoders, giving
$\bm Q_k=\bm w_k\bm w_k^H$ and $\bm Q_s=\bm W_s\bm W_s^H$.
The statistical transmit covariance is $\bm Q=\E[\bm x[n]\bm x^H[n]]=\bm W\bm W^H$,
with
\begin{equation}\label{eq:covariance_decomposition}
    \bm Q
    =
    \sum_{k=1}^{K}\bm Q_k+\bm Q_s,
    \qquad
    \tr(\bm Q)\le P .
\end{equation}
Here, $P>0$ is the total transmit-power budget.

The communication channel of user $k$, denoted by $\bm h_k\in\C^{N_t}$, is
assumed available at the transmitter. It is acquired through uplink training
under channel reciprocity or through CSI feedback. The channels are collected
as $\bm H=[\bm h_1,\ldots,\bm h_K]$.

Target $m$ has an unknown angle $\theta_m$ and complex reflection coefficient
$\beta_m$, with the latter treated as a nuisance parameter in angle estimation.
Past sensing observations or tracking estimates, collected in $z$, determine
their joint posterior. This distribution is held fixed during the current
covariance design. The parameters are collected as
$\bm\theta=[\theta_1,\ldots,\theta_M]^T$ and
$\bm\beta=[\beta_1,\ldots,\beta_M]^T$.
\modelchange{At an $N_r$-element receive array, the sensing observation is modeled as
\begin{equation}\label{eq:sensing_observation}
 \bm y_s[n]=\sum_{m=1}^M\beta_m\bm b(\theta_m)
              \bm a^H(\theta_m)\bm x[n]+\bm n_s[n],
\end{equation}
where $\bm n_s[n]\sim\mathcal{CN}(\bm0,\sigma_s^2\bm I_{N_r})$.
The noise samples are mutually independent and independent of the target
parameters and transmitted signal. The unit-norm transmit steering vector is
given by}
\begin{equation}\label{eq:ula_steering}
 [\bm a(\theta)]_r=N_t^{-1/2}e^{\mathrm j\pi(r-1)\sin\theta},
 \qquad r=1,\ldots,N_t.
\end{equation}
\modelchange{The receive steering vector $\bm b(\theta)$ is defined similarly with
$N_r$ elements. The receive array is held fixed during transmit covariance
optimization.}

\modelchange{The $N_s$ transmitted samples are known to the receiver. Their sample
covariance is $\bm Q_X=N_s^{-1}\sum_n\bm x[n]\bm x^H[n]$.
When the random data are independent of the target state conditional on $z$,
$\E[\bm Q_X\mid z]=\bm Q$. After averaging over this waveform randomness, the
observation FIM depends on $\bm Q$. Controlled waveforms can also realize
$\bm Q_X=\bm Q$ when $N_s\ge\rank(\bm Q)$.}

\subsection{Bayesian Sensing Criterion}

Because each reflection coefficient in~\eqref{eq:sensing_observation} is
complex, the target state is represented by the real vector
\begin{equation}\label{eq:system_target_parameter}
    \bm\xi
    =
    [\theta_1,\ldots,\theta_M,
    \beta_{R,1},\ldots,\beta_{R,M},
    \beta_{I,1},\ldots,\beta_{I,M}]^T ,
\end{equation}
where $\beta_{R,m}=\Re\{\beta_m\}$ and
$\beta_{I,m}=\Im\{\beta_m\}$, so $\bm\xi\in\mathbb R^{3M}$.
Its conditional density is denoted by $p(\bm\xi\mid z)$.
\modelchange{The steering vectors are collected in
$\bm A(\bm\theta)=[\bm a(\theta_1),\ldots,\bm a(\theta_M)]$.}
For each realization, we abbreviate $\bm A(\bm\theta)$ as $\bm A$ and collect
the angle derivatives as
\begin{equation}\label{eq:system_crb_matrices}
    \dot{\bm A} =[\dot{\bm a}_1,\ldots,\dot{\bm a}_M],
\end{equation}
where $\bm a_m=\bm a(\theta_m)$ and
$\dot{\bm a}_m=\partial\bm a(\theta)/\partial\theta|_{\theta=\theta_m}$.
The receive responses are denoted by $\bm b_m=\bm b(\theta_m)$ and
$\dot{\bm b}_m=\partial\bm b(\theta)/\partial\theta|_{\theta=\theta_m}$.

\modelchange{The observation FIM is determined by the sensitivity of the echo channel
to $\bm\xi$. Let $\bm L_m=\bm b_m\bm a_m^H$ and
$\dot{\bm L}_m=\dot{\bm b}_m\bm a_m^H+\bm b_m\dot{\bm a}_m^H$.
The corresponding channel derivatives are
\begin{equation}\label{eq:vector_derivatives}
 \bm D_{\theta_m}=\beta_m\dot{\bm L}_m,\quad
 \bm D_{\beta_{R,m}}=\bm L_m,\quad
 \bm D_{\beta_{I,m}}=\mathrm j\bm L_m.
\end{equation}
After averaging over the waveform, the Gaussian FIM has entries
$[\bm J_\xi^{\rm obs}]_{ij}=\alpha_s\Re\tr(\bm D_i^H\bm D_j\bm Q)$,
where $\alpha_s=2N_s/\sigma_s^2$. The FIM is partitioned into angle information,
reflection-coefficient information, and their coupling as
\begin{equation}\label{eq:sensing_fim_compact}
 \bm J_\xi^{\rm obs}=\alpha_s
 \setlength{\arraycolsep}{1.5pt}
 \begin{bmatrix}
 \Re\bm S_{\theta\theta}&\Re\bm S_{\theta\beta}&-\Im\bm S_{\theta\beta}\\
 (\Re\bm S_{\theta\beta})^T&\Re\bm R_{aa}&-\Im\bm R_{aa}\\
 (-\Im\bm S_{\theta\beta})^T&\Im\bm R_{aa}&\Re\bm R_{aa}
 \end{bmatrix},
\end{equation}
where
\begin{equation}\label{eq:system_fim_blocks}
\begin{aligned}
 [\bm R_{aa}]_{mn}&=\tr(\bm L_m^H\bm L_n\bm Q),\\
 [\bm S_{\theta\theta}]_{mn}
 &=\beta_m^*\beta_n\tr(\dot{\bm L}_m^H\dot{\bm L}_n\bm Q),\\
 [\bm S_{\theta\beta}]_{mn}
 &=\beta_m^*\tr(\dot{\bm L}_m^H\bm L_n\bm Q).
\end{aligned}
\end{equation}
The observation information is therefore linear in $\bm Q$.
It can also be written as
$\bm J_\xi^{\rm obs}=\alpha_s\Re\sum_r\bm\Gamma_r^H\bm Q\bm\Gamma_r$,
where $\bm\Gamma_r=[\bm D_1^H\bm e_r,\ldots,\bm D_{3M}^H\bm e_r]$
and $\bm e_r$ is the $r$th coordinate vector.}

The posterior $p(\bm\xi\mid z)$ obtained from past sensing is used as the prior
for the new observations. Differentiable densities, finite Fisher information,
and vanishing boundary terms are assumed for the Bayesian information
inequality~\cite{VanTrees2001}. With $p=p(\bm\xi\mid z)$, the prior information
and its combination with the averaged observation FIM are
\begin{align}
    \bm J_{\rm prior}(z)
    &=\E[(\nabla_{\bm\xi}\log p)(\nabla_{\bm\xi}\log p)^T\mid z],
    \label{eq:conditional_prior_information}\\
    \bm J_{\rm B}(\bm Q,z)
    &=
    \E[\bm J_\xi^{\rm obs}(\bm Q,\bm\xi)\mid z]
    +
    \bm J_{\rm prior}(z).
    \label{eq:system_bayesian_fim}
\end{align}
The BIM is affine in $\bm Q$ because \eqref{eq:sensing_fim_compact} is linear
and $p(\bm\xi\mid z)$ is fixed.

Let $\bm E_\theta=[\bm I_M,\bm0_{M\times M},\bm0_{M\times M}]
\in\mathbb{R}^{M\times3M}$ select the angle entries of $\bm\xi$. The
corresponding angle-BCRB block is
\begin{equation}\label{eq:system_angle_crb_block}
    \operatorname{BCRB}_{\theta}(\bm Q,z)
    =
    \bm E_\theta
    [\bm J_{\rm B}(\bm Q,z)]^{-1}
    \bm E_\theta^T .
\end{equation}
The joint inverse accounts for the nuisance reflection coefficients.
\modelchange{This block lower-bounds the conditional angle-error second-moment matrix.
The expectation is taken over the target parameters, known random waveform,
and new observations given $z$.}
Its trace bounds the sum of angle mean-square errors and defines the sensing metric
\begin{equation}\label{eq:system_bcrb_scalar}
    \Psi_{\rm B}(\bm Q,z)
    =
    \tr\!\left(
    \operatorname{BCRB}_{\theta}(\bm Q,z)
    \right).
\end{equation}
\modelchange{The prior information and averaged observation FIM are computed from the
same joint posterior. The Gaussian instance and its analytic gain expectation
are specified in Section~\ref{sec:exp_setup}.}

\subsection{Full-Space BCRB Covariance Design}

The signal received by user $k$ is given by
\begin{equation}
    y_k[n]=\bm h_k^H\bm x[n]+n_k[n],
\end{equation}
where $n_k[n]\sim\mathcal{CN}(0,\sigma_k^2)$ is receiver noise with
$\sigma_k^2>0$. Other-user signals and probing are treated as interference.
Define the channel Gram matrix as $\bm R_{h,k}=\bm h_k\bm h_k^H$.
For a prescribed SINR target
$\gamma_k>0$, the resulting requirement is
\begin{equation}\label{eq:system_sinr}
    \frac{\tr(\bm R_{h,k}\bm Q_k)}
    {\sum_{\ell=1,\,\ell\ne k}^{K}\tr(\bm R_{h,k}\bm Q_\ell)
     +\tr(\bm R_{h,k}\bm Q_s)+\sigma_k^2}
    \ge \gamma_k .
\end{equation}
For $k=1,\ldots,K$, cross multiplication converts \eqref{eq:system_sinr}
into the affine margin constraint
\begin{equation}\label{eq:posterior_sinr_affine}
\begin{aligned}
    &c_k(\{\bm Q_\ell\},\bm Q_s)
    =\tr(\bm R_{h,k}\bm Q_k)\\
    &\quad-\gamma_k\Bigl[
    {\textstyle\sum_{\ell=1,\,\ell\ne k}^{K}}\tr(\bm R_{h,k}\bm Q_\ell)\\
    &\qquad\qquad+\tr(\bm R_{h,k}\bm Q_s)+\sigma_k^2\Bigr]
    \ge 0.
\end{aligned}
\end{equation}

The ordinary inverse in~\eqref{eq:system_angle_crb_block} requires
$\bm J_{\rm B}(\bm Q,z)\succ\bm0$. We impose this regularity uniformly over
the feasible covariance domain.
\begin{assumption}[Information regularity]
\label{ass:information_regular}
The posterior instance satisfies
$\lambda_{\min}(\bm J_{\rm B}(\bm Q,z))\ge\mu_J>0$ for every covariance
collection satisfying the positive-semidefinite (PSD), SINR, and total-power
requirements above. This condition holds whenever
$\bm J_{\rm prior}(z)\succeq\mu_J\bm I_{3M}$ because
the observation FIM is positive semidefinite for every $\bm Q\succeq\bm0$.
\end{assumption}

The sensing criterion, SINR, and power depend on precoders only through
covariance blocks. These are optimized with enough probing streams to realize
any $\bm Q_s\succeq\bm0$.
User-block rank-one constraints are relaxed. Under this model, rank-one user
beams can be recovered while preserving the total covariance and every SINR.
Under \textbf{Assumption~\ref{ass:information_regular}}, the Schur complement
gives an exact epigraph representation of the trace-inverse BCRB. The resulting
SDP is

\begin{subequations}\label{prob:system_bcrb_design}
\begin{align}
    \underset{
    \substack{\{\bm Q_k\succeq\bm0\},\,\bm Q_s\succeq\bm0,\\
    \bm Z\succeq\bm0}
    }
    {\operatorname{minimize}}\quad
    & \tr(\bm Z)\\
    \operatorname{subject\ to}\quad
    &
    \begin{bmatrix}
    \bm J_{\rm B}(\bm Q,z) & \bm E_\theta^T\\
    \bm E_\theta & \bm Z
    \end{bmatrix}
    \succeq\bm0,\\
    & c_k(\{\bm Q_\ell\},\bm Q_s)\ge0,\quad \forall k,\\
    & \tr(\bm Q)\le P ,
\end{align}
\end{subequations}
where $\bm Z\in\mathbb R^{M\times M}$ is the PSD epigraph matrix, and
$\bm Q$ is the total covariance in
\eqref{eq:covariance_decomposition}. The linear matrix inequality (LMI)
enforces
$\bm Z\succeq \bm E_\theta[\bm J_{\rm B}(\bm Q,z)]^{-1}\bm E_\theta^T$, and
minimizing $\tr(\bm Z)$ makes the epigraph tight. The linear objective and
affine or positive-semidefinite cone constraints make the formulation convex.

\modelchange{For subsequent analysis, the BIM is written as
$\bm J_{\rm B}(\bm Q,z)=\bm J_{\rm prior}(z)+\mathcal J_z(\bm Q)$,
where $\mathcal J_z$ is the linear observation-information map.
The covariance blocks are indexed by $\mathcal I=\{1,\ldots,K,s\}$.}
\modelchange{The SDP contains $K+1$ covariance blocks of size $N_t\times N_t$.
A full-space solution therefore becomes increasingly
expensive as the array grows. A sufficient low-dimensional basis is also
difficult to identify a priori.}
We seek a low-dimensional design that satisfies the communication constraints
and admits a computable bound on its full-space optimality gap.

\section{\structurechange{Posterior Subspace Structure}}
\label{sec:theory}

\structurechange{For the full-space covariance SDP in Section~\ref{sec:system}, exact
dimension reduction must preserve the SINR constraints and BCRB objective.
The fixed-response case is considered first. Posterior uncertainty is then
introduced, and the support of an optimal design is bounded.}

\subsection{\structurechange{Exact Projection under Posterior Uncertainty}}

\structurechange{The local sensing FIM is first evaluated at a nominal target state
$\bm\xi_0$, using the corresponding steering vectors and their derivatives.
These are collected in $\bm A_0$ and $\dot{\bm A}_0$. Their joint span with
the user channels satisfies
\[
 \dim\spanop[\bm H,\bm A_0,\dot{\bm A}_0]\le K+2M.
\]
Under this projection, the covariance blocks remain PSD. The SINR margins
and fixed-response information matrix are preserved without increasing the
transmit power. The same objective value is therefore attained within this
span, as in the fixed-response reduction of~\cite{Fang2025ISACOBS}.
The projection argument is given in Supplementary
\textbf{Proposition~\ref{prop:posterior_response_observability}}.}

\structurechange{Under posterior uncertainty, the steering and derivative responses vary
with the target state. The observation term in the BIM therefore involves
posterior-averaged response products. These products are preserved by
retaining their column and row spaces. The resulting exact task subspace is
specified in Section~\ref{app:subspace_structure} of the supplement.}

\structurechange{We next examine whether this subspace can have dimension below $N_t$.
Using spatial frequency $u=\sin\theta$, write $\bm a(u)$ for the ULA steering
vector. Its Vandermonde structure gives the following result.}

\begin{proposition}[Posterior steering moment is full rank]
\label{prop:posterior_moment_full_rank}
Let the posterior spatial frequency $u$ have conditional density
$p_z(u)=p(u\mid z)>0$ almost everywhere on a nonzero-width interval
$\mathcal U$. Define
\begin{equation}
    \bm R_a
    =
    \E[\bm a(u)\bm a^H(u)\mid z].
\end{equation}
Then $\bm R_a\succ\bm0$. If an orthogonal projector $\bm P$ preserves the
posterior-averaged steering-response energy
\begin{equation}
    \tr(\bm R_a\bm Q)
    =
    \tr(\bm R_a\bm P\bm Q\bm P),
    \qquad \forall\,\bm Q\succeq\bm0 ,
\end{equation}
then $\bm P=\bm I$.
\end{proposition}

\begin{IEEEproof}
See the supplementary material.
\end{IEEEproof}

\structurechange{Since the receive steering vectors have unit norm,
$\tr(\bm R_a\bm Q)$ appears in the posterior average of the gain-information
block in \eqref{eq:sensing_fim_compact}.
By \textbf{Proposition~\ref{prop:posterior_moment_full_rank}}, preserving this
information for every covariance requires the full-space projector.
Exact preservation therefore requires retaining $N_t\times N_t$ covariance blocks.}

\subsection{\structurechange{Optimal Low-Dimensional Support}}

\structurechange{For covariance optimization, it is sufficient to preserve one optimal
design. Its support can be bounded using the beam-count theorem of Attiah
and Yu~\cite[Thm.~3]{AttiahYu2026HowMany}.
Applying this result to the information map $\mathcal J_z$ gives the
following proposition.}

\structurechange{\begin{proposition}[Optimal-support existence]
\label{prop:optimal_support}
Write $\bm J_{\rm B}=\bm J_{\rm prior}+\mathcal J_z(\bm Q)$.
Suppose \eqref{prob:system_bcrb_design} is strictly feasible and
\textbf{Assumption~\ref{ass:information_regular}} holds. Define
$d_J=\dim_{\mathbb R}\range(\mathcal J_z)$. There exists an optimal covariance
collection whose total covariance satisfies
\begin{equation}
 q_{\rm opt}=\rank\Bigl(\sum_{i\in\mathcal I}\bm Q_i^\star\Bigr)
 \le\left\lfloor\sqrt{K^2+d_J}\right\rfloor.
 \label{eq:optimal_support_bound}
\end{equation}
An optimality-preserving subspace containing $\range(\bm H)$ therefore exists
with dimension at most $\min\{N_t,\rank(\bm H)+q_{\rm opt}\}$.
\end{proposition}}

\structurechange{\begin{IEEEproof}
The covariance feasible set is compact. Under
\textbf{Assumption~\ref{ass:information_regular}}, the objective is continuous,
so an optimum is attained. Rank-one user beams and a probing covariance
realize this optimum without changing the total covariance or any SINR.
Since $\mathcal J_z$ is determined by $d_J$ real linear functionals of
$\bm Q$, the angle-BCRB is a $d_J$-quadratic metric in the beamformers.
Probing is treated as interference at every user, so the case without
interference cancellation in~\cite[Thm.~3]{AttiahYu2026HowMany} applies.
The same objective value is therefore attained using at most
$\lfloor\sqrt{K^2+d_J}\rfloor$ beamformers under the SINR and power constraints.
The resulting covariance rank is bounded by the number of beamformers.
Adjoining $\range(\bm H)$ gives the stated subspace dimension.
\end{IEEEproof}}

\structurechange{Following~\cite[Eq.~(77)]{AttiahYu2026HowMany}, $d_J$ is bounded by counting
the independent real entries in \eqref{eq:sensing_fim_compact}.
The real symmetric angle block contributes at most $M(M+1)/2$ independent
entries. The angle--gain block contributes
at most $2M^2$ real entries. The gain block is determined by an $M\times M$
Hermitian matrix and contributes at most $M^2$. Consequently,
\[
 d_J\le M(M+1)/2+2M^2+M^2=(7M^2+M)/2.
\]
For fixed $K$ and $M$, the resulting support bound is independent of $N_t$.
For $K=M=2$, an optimal support of at most four dimensions exists.
Including the channel span gives a subspace of at most six dimensions.
These dimensions apply to an optimum of the current posterior instance.
A larger basis may be generated during subspace expansion.}

\structurechange{Exact preservation of the posterior information map can require the full
antenna space. The optimal-support existence bound is instead controlled by
the user and target counts. These results motivate a search for a subspace
that meets a prescribed full-space objective gap.}

\section{Certified Dynamic Subspace Optimization}
\label{sec:reduced_isac_optimization}

\methodchange{Further improvement of a feasible design may require a more accurate
reduced solve or additional antenna directions. The reduced- and full-space
gap bounds distinguish these cases. When the reduced solve is sufficiently
accurate but the full-space bound exceeds tolerance, an omitted direction
is identified from the covariance dual slacks. The current feasible design
is retained during expansion. The initial basis, restricted problem, and
certificates are developed below, following the procedure in
Fig.~\ref{fig:system_overview}(b).}

\subsection{\methodchange{Communication-Feasible Fisher Initialization}}
\label{sec:posterior_isac_covariances}

\methodchange{The channel span is retained in the initial basis to preserve
communication feasibility. Its channel basis is}
\begin{equation}
    \bm U_{\rm c}
    =
    \operatorname{orth}(\bm H),
    \qquad
    q_{\rm c}=\rank(\bm H).
\end{equation}
\methodchange{When a basis contains $\bm U_{\rm c}$, projection preserves every SINR
term without increasing power.}

\methodchange{Sensing directions are ranked at a reference covariance
$\bm Q_{\rm ref}\succeq0$ with $\bm J_{\rm B}(\bm Q_{\rm ref},z)\succ0$.
The reference need not satisfy the SINR constraints.
An isotropic covariance is used in the experiments.
Omitting the reference arguments, define}
\begin{equation}\label{eq:bcrb_sensitivity}
\begin{aligned}
    \bm S_{\rm B}
    &=
    \bm J_{\rm B}^{-1}
    \bm E_\theta^T\bm E_\theta
    \bm J_{\rm B}^{-1},\\
    d\Psi_{\rm B}
    &=
    -\tr(\bm S_{\rm B}\,d\bm J_{\rm B}).
\end{aligned}
\end{equation}
\methodchange{Let $\mathcal J_z^\ast$ denote the real-trace adjoint of the
observation-information map in Section~\ref{sec:system}. Through this adjoint,
the weighting gives the antenna-space sensitivity matrix}
\begin{equation}\label{eq:fisher_adjoint_sensitivity}
    \bm T_{\rm F}(\bm Q_{\rm ref},z)
    =
    \mathcal J_z^\ast(\bm S_{\rm B}(\bm Q_{\rm ref},z))\succeq\bm0.
\end{equation}
For any unit-norm direction $\bm v$ and sufficiently small $\tau>0$,
\begin{equation}\label{eq:fisher_first_order}
\begin{aligned}
    &\Psi_{\rm B}(\bm Q_{\rm ref}+\tau\bm v\bm v^H,z)\\
    &\quad=\Psi_{\rm B}(\bm Q_{\rm ref},z)
    -\tau\bm v^H\bm T_{\rm F}(\bm Q_{\rm ref},z)\bm v
    +O(\tau^2).
\end{aligned}
\end{equation}
Thus, the dominant eigenvectors of $\bm T_{\rm F}$ identify the sensing
directions with the largest first-order BCRB decrease per unit added power.
The supplementary material gives the response-moment representation, proves
positive semidefiniteness, and records the corresponding Ky Fan
argument~\cite{GolubVanLoan2013}.

To select sensing directions orthogonal to the retained channel span,
the sensitivity matrix is projected as
\begin{equation}
    \bm P_{\rm c}^{\perp}
    =
    \bm I-\bm U_{\rm c}\bm U_{\rm c}^H,
    \qquad
    \bm T_{{\rm F},\perp}
    =
    \bm P_{\rm c}^{\perp}
    \bm T_{\rm F}(\bm Q_{\rm ref},z)
    \bm P_{\rm c}^{\perp}.
\end{equation}
For a sensing rank $q_{\rm F}$, set
\begin{equation}
    \bm U_{\rm F}
    =
    \operatorname{eig}_{q_{\rm F}}(\bm T_{{\rm F},\perp}),
    \qquad
    \bm U^{(0)}
    =
    \operatorname{orth}([\bm U_{\rm c},\bm U_{\rm F}]).
\end{equation}
Here, $\operatorname{eig}_{q_{\rm F}}(\bm T_{{\rm F},\perp})$ is chosen in
$\range(\bm P_{\rm c}^{\perp})$ and spans a dominant $q_{\rm F}$-dimensional
invariant subspace of the restriction to that range.
\methodchange{For large-array initialization, these eigenvectors are computed by
matrix-free Lanczos iterations. Products with $\bm T_{{\rm F},\perp}$ are
evaluated directly from the posterior response factors, so the sensitivity
matrix need not be assembled explicitly.}

For $q_{\rm F}\le N_t-q_{\rm c}$, the two basis blocks are orthogonal and
$q_0=\rank([\bm U_{\rm c},\bm U_{\rm F}])=q_{\rm c}+q_{\rm F}$.
\methodchange{After power reoptimization, basis sufficiency is assessed by the
full-space certificate. Supplementary Table~\ref{tab:supp_initialization}
compares the initializers under the same expansion mechanism.}

\subsection{\methodchange{Restricted SDP and Feasible Incumbents}}

To reduce the cost of the full-space SDP, all covariance blocks are restricted
to a $q$-dimensional candidate transmit subspace. The original objective and
constraints are evaluated without approximation on the lifted covariances, so
the restriction changes only the available antenna directions. Let
$\bm U\in\C^{N_t\times q}$ be a semi-unitary basis for the candidate
subspace, and let $\bar{\bm Q}_k\succeq\bm0$ and
$\bar{\bm Q}_s\succeq\bm0$ denote the $q\times q$ reduced covariance
variables. Their full-space representations are denoted by
\begin{equation}
    \bm Q_k=\bm U\bar{\bm Q}_k\bm U^H,\qquad
    \bm Q_s=\bm U\bar{\bm Q}_s\bm U^H,
\end{equation}
where the map from reduced coordinates to antenna space is called lifting.
The total reduced covariance is
\begin{equation}
    \bar{\bm Q}
    =
    \sum_{k=1}^{K}\bar{\bm Q}_k+\bar{\bm Q}_s .
\end{equation}
The total transmit covariance is $\bm Q=\bm U\bar{\bm Q}\bm U^H$.
Equivalently, each response moment $\bm R$ is compressed to
$\bm U^H\bm R\bm U$. The resulting BIM is
\begin{equation}
    \bar{\bm J}_{\rm B}(\bar{\bm Q},z)
    =
    \bm J_{\rm B}(\bm U\bar{\bm Q}\bm U^H,z).
\end{equation}
The reduced posterior BCRB design is
\begin{subequations}\label{prob:reduced_bcrb_design}
\begin{align}
    \underset{
    \substack{\{\bar{\bm Q}_k\succeq\bm0\},\,\bar{\bm Q}_s\succeq\bm0,\\
    \bm Z\succeq\bm0}
    }
    {\operatorname{minimize}}\quad
    & \tr(\bm Z)\\
    \operatorname{subject\ to}\quad
    &
    \begin{bmatrix}
    \bar{\bm J}_{\rm B}(\bar{\bm Q},z) & \bm E_\theta^T\\
    \bm E_\theta & \bm Z
    \end{bmatrix}
    \succeq\bm0,\\
    &
    c_k^{\bm U}(\{\bar{\bm Q}_\ell\},\bar{\bm Q}_s)
    \ge0,\quad \forall k,\\
    & \tr(\bar{\bm Q})\le P ,
\end{align}
\end{subequations}
where $c_k^{\bm U}$ is the affine SINR margin in
\eqref{eq:posterior_sinr_affine} with each Gram matrix $\bm R_{h,k}$ replaced
by $\bm U^H\bm R_{h,k}\bm U$.
Problem~\eqref{prob:reduced_bcrb_design} is the full-space SDP restricted to
$\range(\bm U)$ and remains convex. Any exactly feasible reduced point lifts
to a feasible point of the full-space SDP for any semi-unitary $\bm U$.
Since solver feasibility is approximate, numerical candidates are checked
against the original constraints.
\methodchange{The following result supplies a feasible reduced point.}

\methodchange{\begin{lemma}[Reduced-space feasibility]
\label{lem:base_feasibility}
Let $\bm U_0$ be semi-unitary and contain $\spanop(\bm H)$, with
$\bm P_0=\bm U_0\bm U_0^H$. For PSD blocks satisfying all SINR and
power constraints, define
\[
 \widehat{\bm Q}_i=\bm P_0\bm Q_i\bm P_0,\qquad
 \widehat{\bar{\bm Q}}_i=\bm U_0^H\widehat{\bm Q}_i\bm U_0.
\]
If $\widehat{\bm J}_{\rm B}=\bm J_{\rm B}(\sum_i\widehat{\bm Q}_i,z)\succ0$,
then the projected blocks retain SINR and power feasibility. Together with
$\widehat{\bm Z}=\bm E_\theta\widehat{\bm J}_{\rm B}^{-1}\bm E_\theta^T$,
they give a feasible point of \eqref{prob:reduced_bcrb_design}.
\end{lemma}}

\begin{IEEEproof}
See the supplementary material.
\end{IEEEproof}

\methodchange{\begin{remark}[Constructing a feasible reduced point]
\label{rem:feasible_initialization}
Communication feasibility can be sought in $\spanop(\bm H)$.
A positive-definite prior then ensures BIM regularity. Any feasible reduced
point can initialize the exact algorithm. A strictly feasible point also
provides a reference for numerical restoration.
\end{remark}}

Feasibility alone does not show that the candidate subspace is sufficient for
optimization. To enlarge the search space, we append orthonormal directions
collected in $\bm V$ to the current basis.
Orthonormalization gives
\begin{equation}
    \bm U_+=\operatorname{orth}([\bm U,\bm V]),
    \qquad
    \range(\bm U)\subseteq\range(\bm U_+).
\end{equation}
The following result shows that this enlargement retains every feasible design
already found in $\range(\bm U)$.

\begin{lemma}[Feasible expansion]
\label{lem:feasible_expansion}
Let $(\{\bar{\bm Q}_i\}_{i\in\mathcal I},\bm Z)$ be feasible for
\eqref{prob:reduced_bcrb_design} with basis $\bm U$, and set
$\bm Q_i=\bm U\bar{\bm Q}_i\bm U^H$. For any semi-unitary $\bm U_+$ satisfying
$\range(\bm U)\subseteq\range(\bm U_+)$, define
\begin{equation}
    \bar{\bm Q}_{i,+}
    =
    \bm U_+^H\bm Q_i\bm U_+,
    \qquad i\in\mathcal I .
\end{equation}
Then
\begin{equation}
    \bm U_+\bar{\bm Q}_{i,+}\bm U_+^H=\bm Q_i,\qquad i\in\mathcal I .
\end{equation}
With the same epigraph block $\bm Z$,
$(\{\bar{\bm Q}_{i,+}\}_{i\in\mathcal I},\bm Z)$ is feasible for
\eqref{prob:reduced_bcrb_design} with basis $\bm U_+$ and has the same objective
value.
\end{lemma}

\begin{IEEEproof}
See the supplementary material.
\end{IEEEproof}

Thus, expansion never discards the current feasible design or worsens its
objective value.

\subsubsection{\methodchange{Restoring a Valid Primal Upper Bound}}
\label{sec:primal_restoration}
\methodchange{A valid upper bound is obtained by restoring primal feasibility before
evaluating the objective. Each reduced covariance is first projected onto
the PSD cone with a positive numerical margin.
Let $\{\bm X_i\}$ denote these blocks and let
$\{\bm R_i\}$ be a reduced PSD reference whose affine SINR and power slacks
$s_j(\bm R)$ are strictly positive. With $s_j$ denoting all $K+1$ affine
slacks and safety margins $0<\tau_j<s_j(\bm R)$, set
\begin{equation}
 t=\max_j\left\{\frac{[\tau_j-s_j(\bm X)]_+}
 {s_j(\bm R)-s_j(\bm X)}:\ s_j(\bm X)<\tau_j\right\},
 \label{eq:primal_mixing}
\end{equation}
where $[x]_+=\max\{x,0\}$ and $t=0$ when the set is empty. Use
$\bar{\bm Q}_i=(1-t)\bm X_i+t\bm R_i$.
Affineness gives $s_j(\bar{\bm Q})\ge\tau_j$ and convexity preserves PSD.
The BIM and tight Schur epigraph are then recomputed. For full-column-rank
channels, zero-forcing beams in $\range(\bm H)$ supply the reference whenever
their minimum required power is strictly below $P$. In other cases, a
strictly feasible reference must be found separately for this restoration
step.

A factorized lifted covariance is used in floating-point arithmetic.
Positive reduced PSD and affine margins are verified, and the objective is
inflated by an inverse-solve residual bound. Error allowances and restoration
changes are reported in the supplement.}

\subsection{\methodchange{Full-Space Bounds from Reduced Dual Variables}}

The reduced SDP considers only directions within $\range(\bm U)$. The
full-space covariance dual slacks are therefore used to check omitted
directions and obtain a lower bound on the full-space optimum.

\methodchange{To construct the slack of each covariance block $i\in\mathcal I$, the
coefficients of the SINR margin
\eqref{eq:posterior_sinr_affine} are collected as}
\begin{equation}\label{eq:sinr_blockwise_form}
\begin{aligned}
    c_k(\{\bm Q_\ell\},\bm Q_s)
    &=
    \sum_{i\in\mathcal I}\tr(\bm C_{k,i}\bm Q_i)
    -\gamma_k\sigma_k^2,
    \end{aligned}
\end{equation}
where
\begin{equation}
\begin{aligned}
    \bm C_{k,i}
    &=
    \begin{cases}
    \bm R_{h,k}, & i=k,\\
    -\gamma_k\bm R_{h,k}, & i\ne k .
    \end{cases}
\end{aligned}
\end{equation}
The probing covariance corresponds to $i=s$ and therefore uses the second
coefficient.

The reduced Schur, SINR, and total-power constraints are the original
constraints evaluated at the lifted covariances. Their dimensions are
unchanged, so the corresponding multipliers can be reused as candidate
full-space dual variables. Let $\bm Y$ denote the dual matrix of the Schur LMI in
\eqref{prob:system_bcrb_design}, partitioned conformably as
\begin{equation}
    \bm Y
    =
    \begin{bmatrix}
    \bm Y_J & \bm Y_{J\theta}\\
    \bm Y_{\theta J} & \bm Y_Z
    \end{bmatrix}.
\end{equation}
Let $\mu_k$ and $\nu$ denote the multipliers of the $k$th SINR constraint and
the total-power constraint, respectively. A dual candidate
$(\bm Y,\bm\mu,\nu)$ is admissible when
\begin{equation}
    \bm Y\succeq\bm0,\qquad
    \bm I-\bm Y_Z\succeq\bm0,\qquad
    \bm\mu\ge\bm0,\qquad
    \nu\ge0 .
\end{equation}
These signs follow from the corresponding inequalities, while
$\bm I-\bm Y_Z\succeq\bm0$ bounds the Lagrangian below over $\bm Z\succeq\bm0$.
PSD projection, uniform rescaling of $\bm Y$, and nonnegative scalar clipping
enforce these conditions, as detailed in the supplementary material.
Covariance-block conditions are checked below.

\methodchange{The adjoint $\mathcal J_z^\ast$ maps the information-matrix coefficient
$\bm Y_J$ back to the covariance domain. For every Hermitian perturbation
$d\bm Q$, it satisfies}
\begin{equation}
    \tr(\bm Y_J\mathcal J_z(d\bm Q))
    =
    \tr(\mathcal J_z^\ast(\bm Y_J)\,d\bm Q).
\end{equation}
The conditions $\bm Q_i\succeq\bm0$ and $\bm Z\succeq\bm0$ are kept in the
conic domain. The Schur, SINR, and total-power constraints are dualized with
the multipliers above, giving the blockwise Lagrangian
\begin{equation}\label{eq:dual_lagrangian_decomposition}
    \mathcal L
    =
    d_0
    +
    \tr((\bm I-\bm Y_Z)\bm Z)
    +
    \sum_{i\in\mathcal I}\tr(\bm G_i\bm Q_i),
\end{equation}
where the constant term is
\begin{align}\label{eq:explicit_dual_lower_constant}
    d_0(\bm Y,\bm\mu,\nu)
    &=
    -\tr(\bm Y_J\bm J_{\rm prior}(z))
    -
    2\Re\{\tr(\bm Y_{J\theta}^H\bm E_\theta^T)\} \notag\\
    &\quad
    +
    \sum_{k=1}^{K}\mu_k\gamma_k\sigma_k^2
    -
    \nu P .
\end{align}
The coefficient of covariance block $\bm Q_i$ is its dual slack
\begin{equation}\label{eq:block_lagrangian_gradient}
    \bm G_i
    =
    -\mathcal J_z^\ast(\bm Y_J)
    +
    \nu\bm I
    -
    \sum_{k=1}^{K}\mu_k\bm C_{k,i}.
\end{equation}
Both $d_0$ and $\{\bm G_i\}$ are formed from the current dual candidate.

\methodchange{At an exact restricted primal--dual optimum, Schur complementarity gives
$\bm Y_J=\bm S_{\rm B}(\bm Q^\star,z)$, where $\bm Q^\star$ is the lifted
total covariance. Hence $\mathcal J_z^\ast(\bm Y_J)=\bm T_{\rm F}(\bm Q^\star,z)$.
The slacks combine this BCRB sensitivity with the optimized SINR and power
multipliers, as in the sensing-direction representation
of~\cite{AttiahYu2025Duality}. The initializer evaluates the same sensitivity
at the reference covariance.}

By \eqref{eq:dual_lagrangian_decomposition}, an admissible dual candidate is
feasible for the full dual problem if
$\bm G_i\succeq\bm0$ for every $i\in\mathcal I$. At a full-space
primal--dual optimum, these conditions are accompanied by complementary
slackness
\begin{equation}
    \bm G_i\succeq\bm0,\qquad
    \tr(\bm G_i\bm Q_i)=0,
    \qquad i\in\mathcal I ,
\end{equation}
whereas the restricted dual requires only
$\bm U^H\bm G_i\bm U\succeq\bm0$. This difference permits a direct test for
directions omitted by the current subspace.

Full-space slack feasibility is equivalent to
$\bm v^H\bm G_i\bm v\ge0$ for every unit-norm direction $\bm v$. By the
Rayleigh quotient,
$\min_{\|\bm v\|_2=1}\bm v^H\bm G_i\bm v=\lambda_{\min}(\bm G_i)$.
A negative minimum eigenvalue therefore identifies a violated full-space dual
direction, and a corresponding eigenvector gives its strongest rank-one
violation for the current dual candidate. The worst violation over all covariance
blocks is
\begin{equation}\label{eq:full_gradient_defect}
    \delta_{\rm full}
    =
    \max_{i\in\mathcal I}
    \left[-\lambda_{\min}(\bm G_i)\right]_+ .
\end{equation}
\methodchange{Applying the same spectral test within $\range(\bm U)$ gives the
reduced-subspace defect}
\begin{equation}\label{eq:reduced_dual_slack_defect}
    \eta_{\bm U}
    =
    \max_{i\in\mathcal I}
    \left[
    -\lambda_{\min}(\bm U^H\bm G_i\bm U)
    \right]_+ .
\end{equation}
Since the unit vectors in $\range(\bm U)$ form a subset of the full-space
unit vectors, $0\le\eta_{\bm U}\le\delta_{\rm full}$. The full-space defect
also gives the smallest common shift that makes every covariance slack
positive semidefinite,
\begin{equation}
    \delta_{\rm full}
    =
    \min\{\zeta\ge0:\ \bm G_i+\zeta\bm I\succeq\bm0,\ \forall i\in\mathcal I\}.
    \label{eq:minimal_common_dual_shift}
\end{equation}
For an admissible dual candidate, each slack contains $\nu\bm I$, so increasing $\nu$ by $\zeta$ adds
$\zeta\bm I$ to every slack. The smallest full-space
repair is therefore $\nu^+=\nu+\delta_{\rm full}$, which changes $d_0$ by
$-P\delta_{\rm full}$. Together with admissibility, the repaired slacks ensure
full dual feasibility, so weak duality gives the lower bound
$d_0-P\delta_{\rm full}$. With $(\bm Y,\bm\mu)$ fixed, this is the largest
lower bound obtainable by a nonnegative increase of the current $\nu$.
Replacing $\delta_{\rm full}$ by $\eta_{\bm U}$
gives the restricted-space lower bound $d_0-P\eta_{\bm U}$.

\methodchange{\begin{remark}[Eliminating the power multiplier]
\label{rem:optimal_power_multiplier}
For fixed $(\bm Y,\bm\mu)$, let
$\bm B_i=\mathcal J_z^\ast(\bm Y_J)+\sum_k\mu_k\bm C_{k,i}$.
The best feasible power multipliers are
\begin{align}
 \nu_{\rm full}^\star&=\max\{0,\max_i\lambda_{\max}(\bm B_i)\},\\
 \nu_{\bm U}^\star&=\max\{0,\max_i\lambda_{\max}(\bm U^H\bm B_i\bm U)\}.
\end{align}
Indeed, dual feasibility requires $\nu\ge\nu_\bullet^\star$, and the dual
objective decreases with $\nu$. Evaluating the preceding scalar-repair
formulas from the admissible dual candidate $(\bm Y,\bm\mu,0)$ gives exactly these
optimal multipliers. The implementation therefore uses $\nu=0$ before
repair, with conservative spectral enclosures used for numerical lower bounds.
\end{remark}}

Computable primal--dual gaps are obtained by comparing the repaired lower
bounds with the objective of a lifted feasible primal point. The feasible
design with the smallest objective value found so far is retained as the
incumbent. For any such point
$(\{\widehat{\bm Q}_i\}_{i\in\mathcal I},\widehat{\bm Z})$, define
\begin{align}
    \Delta_{\rm red}
    &=
    \tr(\widehat{\bm Z})
    -
    d_0(\bm Y,\bm\mu,\nu)
    +
    P\eta_{\bm U},\\
    \Delta_{\rm full}
    &=
    \tr(\widehat{\bm Z})
    -
    d_0(\bm Y,\bm\mu,\nu)
    +
    P\delta_{\rm full}.
    \label{eq:certificate_gaps}
\end{align}
\methodchange{The defects depend on the power multiplier of the dual candidate.
With the implemented choice $\nu=0$, $\eta_{\bm U}$ can remain positive at an
exact restricted optimum. Inner-solve accuracy is therefore assessed by
$\Delta_{\rm red}$.}
The gap difference
$\Delta_{\rm full}-\Delta_{\rm red}
=P(\delta_{\rm full}-\eta_{\bm U})$ is the certificate-extension term. It is
the additional repair cost incurred when the same dual candidate is extended from
the reduced subspace to the full antenna space. This repair cost is not, in
general, the optimal-value loss caused by the subspace restriction.

\subsection{Full-Space Certificate and Expansion Direction}

The following theorem shows that these gaps bound the reduced- and full-space
suboptimality for any feasible lifted point and admissible dual candidate.

\begin{theorem}[Scalar-repaired full-space certificate]
\label{thm:dynamic_certificate}
Consider any subspace $\range(\bm U)$ and a candidate obtained after lifting
its reduced covariance variables. Let
$(\{\widehat{\bm Q}_i\}_{i\in\mathcal I},\widehat{\bm Z})$ denote the lifted
primal point and assume that it is feasible for the full SDP
\eqref{prob:system_bcrb_design}. Let
$(\widehat{\bm Y},\widehat{\bm\mu},\widehat\nu)$ be an admissible dual candidate,
and form $d_0$, $\{\bm G_i\}$, $\eta_{\bm U}$, $\delta_{\rm full}$,
$\Delta_{\rm red}$, and $\Delta_{\rm full}$ as in
\eqref{eq:explicit_dual_lower_constant}--\eqref{eq:certificate_gaps}.
Let $p^\star$ be the optimum value of the full SDP, and let
$p_{\bm U}^\star$ be the optimum value of the same SDP restricted to
$\range(\bm U)$. Then
\begin{align}
    0
    &\le
    \tr(\widehat{\bm Z})-p_{\bm U}^\star
    \le
    \Delta_{\rm red},\\
    0
    &\le
    \tr(\widehat{\bm Z})-p^\star
    \le
    \Delta_{\rm full}.
    \label{eq:dynamic_gap_certificate}
\end{align}
Moreover,
\begin{equation}
    \Delta_{\rm full}
    =
    \Delta_{\rm red}
    +
    P(\delta_{\rm full}-\eta_{\bm U}),
\end{equation}
where the second term is the certificate-extension term.
\end{theorem}

\begin{IEEEproof}
\methodchange{Increasing $\widehat\nu$ by $\delta_{\rm full}$ makes every covariance slack
PSD while retaining the other dual conditions. Weak duality therefore gives
$p^\star\ge d_0-P\delta_{\rm full}$. The feasible incumbent supplies
$p^\star\le\tr(\widehat{\bm Z})$. Repeating the argument on the restricted
slacks with shift $\eta_{\bm U}$ proves the reduced bound. Subtraction gives
the decomposition. The supplement provides the full Lagrangian argument.}
\end{IEEEproof}

For the stopping test, these repaired primal--dual gaps are normalized by
the positive feasible objective value,
\begin{equation}\label{eq:normalized_certificates}
    \widehat\Delta_{\rm red}
    =
    \frac{\Delta_{\rm red}}{\tr(\widehat{\bm Z})},
    \qquad
    \widehat\Delta_{\rm full}
    =
    \frac{\Delta_{\rm full}}{\tr(\widehat{\bm Z})}.
\end{equation}
Let $0<\varepsilon_{\rm in}<\varepsilon_{\rm out}$ denote the inner tolerance
for the reduced solve and the outer tolerance for full-space certification,
respectively.
Because both gaps use the same positive normalization,
$\widehat\Delta_{\rm red}\le\varepsilon_{\rm in}<\varepsilon_{\rm out}
<\widehat\Delta_{\rm full}$ implies
$\delta_{\rm full}>\eta_{\bm U}$. The reduced solution is then accurate enough
for the restricted problem but not yet certified for the full-space problem,
so the algorithm seeks an outside direction in the full-space slacks.
\methodchange{Accordingly, the algorithm terminates when
$\widehat\Delta_{\rm full}\le\varepsilon_{\rm out}$. Otherwise, the restricted
solve is continued when $\widehat\Delta_{\rm red}>\varepsilon_{\rm in}$,
and the subspace is expanded when
$\widehat\Delta_{\rm red}\le\varepsilon_{\rm in}$.
These decisions are summarized in Algorithm~\ref{alg:dynamic_subspace_expansion}.}

The eigenvectors associated with negative eigenvalues of the full-space slacks
identify such directions. The lemma below shows that, under the preceding
condition, a most-negative eigenvector has a nonzero component outside
$\range(\bm U)$. For an eigenvector $\bm v_i$ of $\bm G_i$, define this
normalized component as
\begin{equation}\label{eq:outside_negative_eigenvector}
    \bm d_i
    =
    \frac{(\bm I-\bm U\bm U^H)\bm v_i}
    {\|(\bm I-\bm U\bm U^H)\bm v_i\|_2},
\end{equation}
whenever the denominator is nonzero.

\begin{lemma}[Expansion outside the current subspace]
\label{lem:outside_direction}
Suppose $\delta_{\rm full}>\eta_{\bm U}$, which holds in particular whenever
$\widehat\Delta_{\rm red}\le\varepsilon_{\rm in}<\varepsilon_{\rm out}
<\widehat\Delta_{\rm full}$. Let $i^\star$ attain $\delta_{\rm full}$ and let
$\bm v^\star$ be any \methodchange{unit-norm} eigenvector with
$\bm G_{i^\star}\bm v^\star=-\delta_{\rm full}\bm v^\star$. Then
$(\bm I-\bm U\bm U^H)\bm v^\star\ne\bm0$. Moreover, appending the normalized
outside residual $\bm d^\star$ of $\bm v^\star$ from
\eqref{eq:outside_negative_eigenvector} gives
$\bm U_+=\operatorname{orth}([\bm U,\bm d^\star])$ with
\begin{equation}\label{eq:violation_exposure}
    \lambda_{\min}(\bm U_+^H\bm G_{i^\star}\bm U_+)
    \le
    -\delta_{\rm full}.
\end{equation}
\end{lemma}

\begin{IEEEproof}
See the supplementary material.
\end{IEEEproof}

\methodchange{The appended residual exposes the current dual violation in the enlarged
subspace. For minimum-power beamforming with SINR constraints, dual
uniqueness is established in~\cite{HuangPalomar2010Dual}.
In BCRB design, the Schur, SINR, and power multipliers are coupled.
Their joint uniqueness is established below under a strict
communication-power margin and is then used to prove strict improvement.
Let $P_{\rm c}(\bm U)$ denote the minimum total power needed to
satisfy the SINR constraints with PSD covariance blocks supported on
$\range(\bm U)$. The condition $P_{\rm c}(\bm U)<P$ provides the strict
reference used in Section~\ref{sec:primal_restoration}.
The sensing-optimal design may still use the full power budget.
Since the channel span is retained, the condition holds on every visited
subspace whenever the full-space communication problem is feasible with
power below $P$.}

\methodchange{\begin{theorem}[Dual uniqueness and strict improvement]
\label{thm:strict_expansion}
Suppose $K\ge1$, $P_{\rm c}(\bm U)<P$, and
\textbf{Assumption~\ref{ass:information_regular}} holds. Then the following
statements hold.
\begin{enumerate}
\item[(i)] The restricted SDP
\eqref{prob:reduced_bcrb_design} has a unique optimal dual triple
$(\bm Y,\bm\mu,\nu)$ for the dual of \eqref{prob:reduced_bcrb_design}.
\item[(ii)] Let a feasible restricted primal point and an admissible dual candidate
satisfy $\Delta_{\rm red}=0$. If $\delta_{\rm full}>\eta_{\bm U}$, append
the direction from \textbf{Lemma~\ref{lem:outside_direction}} to obtain
$\bm U_+$. Then
\begin{equation}\label{eq:strict_expansion_improvement}
 p_{\bm U_+}^\star<p_{\bm U}^\star.
\end{equation}
\end{enumerate}
\end{theorem}}

\begin{IEEEproof}
\methodchange{For (i), the Schur multiplier is fixed by complementarity. Positive receiver
noise powers make the user-block SINR matrix strictly diagonally dominant.
Any remaining multiplier ambiguity yields an optimal primal--dual pair for
the minimum communication-power problem at power $P$.
This contradicts $P_{\rm c}(\bm U)<P$ and establishes joint uniqueness.

For (ii), the repair $\widetilde\nu=\nu+\eta_{\bm U}$ gives the unique restricted
dual optimum because $\Delta_{\rm red}=0$. Its slack along the selected
eigenvector satisfies
\[
 (\bm v^\star)^H(\bm G_{i^\star}+\eta_{\bm U}\bm I)\bm v^\star
 =-\delta_{\rm full}+\eta_{\bm U}<0.
\]
Since $\bm v^\star\in\range(\bm U_+)$, this dual optimum is infeasible in
the enlarged space. Equality of the two primal optima would make an attained
enlarged dual optimum also optimal in the old space, contradicting
uniqueness. The full argument is given in the supplement.}
\end{IEEEproof}

\methodchange{The strict decrease concerns exact restricted solves. At positive inner
tolerances, \textbf{Lemma~\ref{lem:feasible_expansion}} ensures that the best
feasible objective is nonincreasing. The dual candidate changes during
reoptimization, so $\delta_{\rm full}$ and the certificate bound may still
increase or decrease between expansions.}

The following result transfers the certificate in
\textbf{Theorem~\ref{thm:dynamic_certificate}} to rank-one user beams while
preserving the total covariance.

\begin{corollary}[Rank-one transfer of the certificate]
\label{cor:rank_one_transfer}
Let $(\{\bm Q_k\},\bm Q_s,\widehat{\bm Z})$ be any feasible
point of \eqref{prob:system_bcrb_design}. Since $\gamma_k>0$ and $\sigma_k^2>0$,
SINR feasibility implies
$\bm h_k^H\bm Q_k\bm h_k\ge\gamma_k\sigma_k^2>0$ for every
$k=1,\ldots,K$. Define
\begin{equation}\label{eq:rank_one_transfer_map}
\begin{aligned}
    \widetilde{\bm w}_k
    &=
    \frac{\bm Q_k\bm h_k}{(\bm h_k^H\bm Q_k\bm h_k)^{1/2}},
    \qquad\\
    \widetilde{\bm Q}_k
    &=
    \widetilde{\bm w}_k\widetilde{\bm w}_k^H,\\
    \widetilde{\bm Q}_s
    &=
    \bm Q_s+\sum\nolimits_k(\bm Q_k-\widetilde{\bm Q}_k).
\end{aligned}
\end{equation}
Then every
$\widetilde{\bm Q}_k$ is rank one,
$\bm Q_k-\widetilde{\bm Q}_k\succeq\bm0$, and the
total covariance, all SINR values, the BCRB objective, and the transmit power
are unchanged. If an admissible dual candidate certifies the original point,
the same certificate \eqref{eq:dynamic_gap_certificate} applies after recovery.
\end{corollary}

\begin{IEEEproof}
See the supplementary material.
\end{IEEEproof}

\subsection{\methodchange{Certified Expansion Algorithm}}

\methodchange{Algorithm~\ref{alg:dynamic_subspace_expansion} combines restricted solves,
certificate checks, and basis expansion to obtain a certified transmit covariance.
Any restricted solver providing a primal candidate and dual variables can be used.
In our implementation,
\eqref{prob:reduced_bcrb_design} is solved through the MOSEK Fusion interface~\cite{MosekFusion11}.
The information matrix is conditioned by an invertible congruence, and complex
Hermitian covariance blocks are represented by equivalent real PSD cones.
The mapping and independent checks of its dual objective against the solver
output are given in the supplementary material.}

\methodchange{Stopping and expansion use the gaps after primal restoration and dual
repair. MOSEK is called with a tighter internal tolerance when the reduced
accuracy gate is not met.}

\methodchange{During expansion, the feasible incumbent with the smallest upper bound
is embedded in the enlarged basis and each restricted model is rebuilt.
Certificates are computed
independently of MOSEK's stopping status. The restored reduced covariance
blocks and their basis are returned.}

\begin{algorithm}[!t]
\caption{\protect\raggedright Certified dynamic subspace optimization}
\label{alg:dynamic_subspace_expansion}
\begin{algorithmic}[1]
\REQUIRE Data of \eqref{prob:system_bcrb_design}, $\bm Q_{\rm ref}$,
$0\le q_{\rm F}\le N_t-\rank(\bm H)$, and
$0<\varepsilon_{\rm in}<\varepsilon_{\rm out}$
\STATE $\bm U_{\rm c}=\operatorname{orth}(\bm H)$ and
$\bm U_{\rm F}=\operatorname{eig}_{q_{\rm F}}(\bm T_{{\rm F},\perp})$ with
$\bm T_{{\rm F},\perp}$ from Section~\ref{sec:posterior_isac_covariances}
\STATE Set $\bm U^{(0)}=\operatorname{orth}([\bm U_{\rm c},\bm U_{\rm F}])$
\STATE Initialize the restricted solver in $\range(\bm U^{(0)})$ and obtain a
feasible incumbent as in \textbf{Remark~\ref{rem:feasible_initialization}}.
Set $\ell=0$
\WHILE{true}
    \STATE Run the restricted solver in $\range(\bm U^{(\ell)})$ for the reduced SDP
    in \eqref{prob:reduced_bcrb_design}
    \STATE \methodchange{Restore a feasible primal point using Section~\ref{sec:primal_restoration},
    retain the best incumbent, and form an admissible dual candidate with $\nu=0$}
    \STATE Form $\{\bm G_i\}_{i\in\mathcal I}$ from
    \eqref{eq:block_lagrangian_gradient} and evaluate
    $\eta_{\bm U}$, $\delta_{\rm full}$, and the normalized certificates
    $\widehat\Delta_{\rm red}$, $\widehat\Delta_{\rm full}$ in
    \eqref{eq:normalized_certificates}
    \IF{$\widehat\Delta_{\rm full}\le\varepsilon_{\rm out}$}
        \STATE \textbf{return} the lifted covariance and its certificate
    \ELSIF{$\widehat\Delta_{\rm red}>\varepsilon_{\rm in}$}
        \STATE Continue the restricted solve in the same subspace
    \ELSE
        \STATE Choose $i^\star\in\mathcal I$ attaining $\delta_{\rm full}$ in
        \eqref{eq:full_gradient_defect} and a most-negative eigenvector
        $\bm v_{i^\star}$ of $\bm G_{i^\star}$
        \STATE $\bm U^{(\ell+1)}\leftarrow
        \operatorname{orth}([\bm U^{(\ell)},\bm d_{i^\star}])$
        \STATE Embed the incumbent via
        \textbf{Lemma~\ref{lem:feasible_expansion}}, warm-start the restricted
        solver when supported, and set
        $\ell\leftarrow\ell+1$
    \ENDIF
\ENDWHILE
\end{algorithmic}
\end{algorithm}

By \textbf{Lemma~\ref{lem:outside_direction}}, each triggered expansion
increases the subspace dimension. The following result bounds the number of
expansions.

\begin{corollary}[Conditional finite termination]
\label{cor:finite_termination}
Suppose that, for every visited fixed subspace, the inner solver supplies after
finitely many inner iterations a full-space feasible primal incumbent and an
admissible dual candidate satisfying
$\widehat\Delta_{\rm red}\le\varepsilon_{\rm in}<\varepsilon_{\rm out}$.
By \textbf{Lemma~\ref{lem:outside_direction}}, every triggered expansion adds
a direction outside the current subspace. Hence,
Algorithm~\ref{alg:dynamic_subspace_expansion} performs at most $N_t-q_0$
expansions and terminates at a feasible incumbent satisfying
$\widehat\Delta_{\rm full}\le\varepsilon_{\rm out}$.
\end{corollary}

\begin{IEEEproof}
See the supplementary material.
\end{IEEEproof}

\subsection{Computational Complexity}\label{sec:computational_complexity}

\methodchange{At dimension $q$, the covariance blocks contain $(K+1)q^2$ real coordinates,
and the Schur epigraph contains $M(M+1)/2$ coordinates. The native real
formulation has $K+1$ PSD cones of order $2q$ and one Schur cone of order
$4M$. Thus subspace restriction reduces the large covariance cones while
leaving the task-sized information structure unchanged. Actual conic cost
also depends on the constraint structure, conditioning, and number of
interior-point iterations. These costs are assessed in the runtime
experiments.}

\methodchange{Initialization and certification both operate in antenna space. For
certification, full Hermitian eigendecompositions are computed with residual
checks, at a cost of $O((K+1)N_t^3)$ per full-space certificate.
Posterior-moment assembly and basis construction are also included in online
time.}

\methodchange{Low-rank factorization $\bm Q_i=\bm L_i\bm L_i^H$ provides another reduced
representation~\cite{Boumal2019BM,Cifuentes2021BM}. The objective and SINR
constraints are nonconvex in the factors. A factorized augmented-Lagrangian
baseline is implemented here with the same primal restoration and full-space
certificate. Its settings are given in the supplement.}

\section{Numerical Experiments}\label{sec:experiments}

Certificate accuracy and posterior sampling are first examined. The dimension
and runtime required by adaptive expansion are then evaluated as the task and
array sizes vary.

\subsection{Simulation Setup}\label{sec:exp_setup}
The half-wavelength ULA uses unit-norm steering vectors and a colocated
receive array with $N_r=N_t$. Unless varied, $K=M=2$, $P=10$,
$\gamma_k=15$ dB, $\sigma_k^2=10^{-3}$, and $\alpha_s=2N_s/\sigma_s^2=1000$.
Target angles have independent Gaussian posteriors with
centers $(-35^\circ,35^\circ)$ and standard deviations $(6^\circ,6^\circ)$.
\modelchange{User channels have Rician factor $3$ and nominal angles
$(-45^\circ,45^\circ)$ with $1^\circ$ uniform jitter.}

\modelchange{The physical gains are mutually independent and independent of the angles,
with $\beta_m\mid z\sim\mathcal{CN}(\bar\beta_m,v_m)$.
The gain parameters are set to $\bar\beta_m=1$ and
$v_m=c_\beta^2|\bar\beta_m|^2$ with $c_\beta=0.5$.
Writing $\sigma_{\theta,m}$ for the angle standard deviation, the prior
information and gain moments are
\begin{align}
 \bm J_{\rm prior}
 &=\diag\!\left(\{\sigma_{\theta,m}^{-2}\}_m,
                 \{2/v_m\}_m,\{2/v_m\}_m\right),\\
 \E[\beta_m^*\beta_n\mid z]
 &=\bar\beta_m^*\bar\beta_n+\mathbf1_{m=n}v_m.
 \label{eq:gain_moments}
\end{align}
By \eqref{eq:gain_moments}, gain uncertainty contributes
$\alpha_s v_m\E[\tr(\dot{\bm L}_m^H\dot{\bm L}_m\bm Q)\mid z]$
to the $m$th angle-diagonal entry.
The angle--gain terms depend on $\bar\beta_m^*$, while the gain block is
independent of the gains. These moments are integrated analytically.
The remaining angle expectation is approximated by Latin hypercube sampling
(LHS), with $S$ draws stratified separately for each target. To account for the aperture and
task size, $S=120\cdot2^{j}$ is used with the smallest $j\ge0$ such that
$S\ge\max\{120,15N_t/16\}(M-1)$. This rule was fixed on separate pilot scenes
and is checked in Section~\ref{sec:exp_validation}.}

The proposed method, denoted Dynamic Fisher, uses
Algorithm~\ref{alg:dynamic_subspace_expansion} with Fisher initialization
and is labeled ``Proposed'' in the figures and tables.
For initialization, $\bm Q_{\rm ref}=(P/N_t)\bm I$ and $q_{\rm F}=M$
are used by default, giving $q_0=K+M$ in the tested settings.
Dynamic Resp-PCA and Dynamic Steering are constructed as initializer ablations.
Resp-PCA denotes response-based principal component analysis.
Unless stated otherwise, conic problems are solved by
native MOSEK on an Apple M4, serially with one thread. A $1\%$ relative
certificate target is used, with a reduced-gap gate one tenth as large.
Restricted solves start at an interior-point tolerance of $10^{-8}$ and
are tightened to $10^{-10}$ when needed. For full-space MOSEK, the initial
tolerance of $10^{-4}$ was selected on separate pilot scenes and is tightened
to $10^{-6}$ and $10^{-8}$ when needed.

The factorized augmented-Lagrangian (AL) baseline uses probing rank two,
also selected on pilot scenes. Both AL and Dynamic Fisher use the matrix-free
initializer of Section~\ref{sec:posterior_isac_covariances}. The static
orderings and calibrated discrete prolate spheroidal sequence (DPSS)
sectors~\cite{Slepian1978} are specified in the supplement.

Main-comparison times include model construction, initialization, all solve
attempts, restoration, and full spectral checks. Runtime ratios are medians
over paired scenes. All methods were timed on the same machine with the same
software and thread configuration.
Input generation and pairing are specified in Supplementary
Section~\ref{app:additional_experiments}.

\subsection{Certificate and Sampling Validation}\label{sec:exp_validation}
\begin{figure}[!t]
 \centering
 \includegraphics[width=\columnwidth]{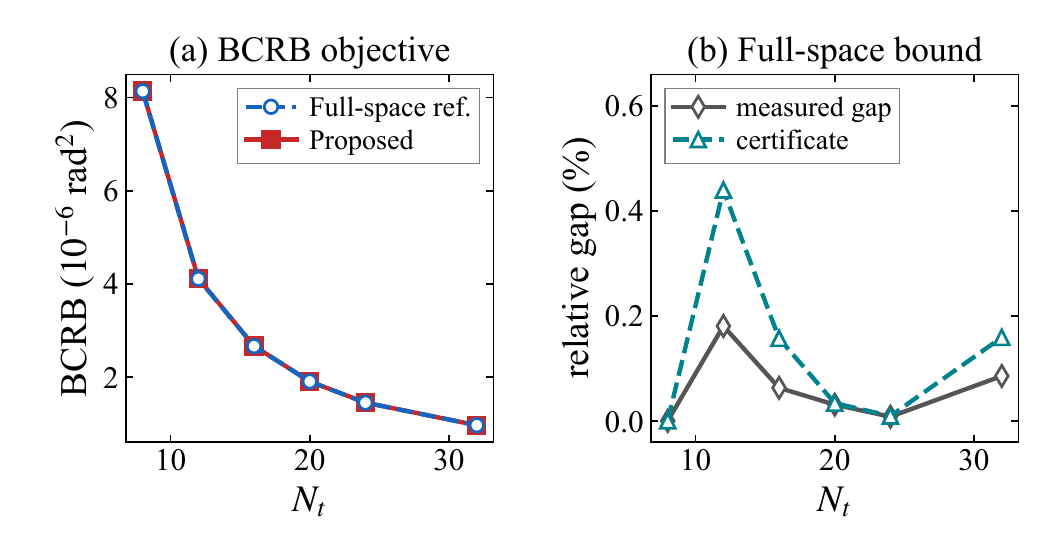}
 \caption{Full-space validation. (a) BCRB. (b) Gaps normalized by $f_{\rm up}$.}
 \label{fig:exp_full_validation}
\end{figure}

The gap bound is checked in Fig.~\ref{fig:exp_full_validation} by comparing
restored designs with native MOSEK full-space references for $N_t=8$--$32$.
The dynamic designs use $q=\RevValidationQRange$, with a largest relative
certificate of $\RevValidationGapMax\%$. Every measured reference gap lies
below its bound.
One paired instance is used at each array size. At $N_t=64$--$512$, the paired
full-space solves used in the runtime comparison bracket the optimum
within $\ReceiveMainReferenceCertMax\%$. In every case, the certificate exceeds the
resulting lower estimate of the true gap, and its median ratio to the upper
estimate $f_{\rm up}-L_{\rm ref}$ is $\ReceiveTightMin$--$\ReceiveTightMax$
across array sizes. Here $f_{\rm up}$ and $L_{\rm ref}$ denote the restored
BCRB upper bound and paired full-space lower bound in physical units.
Saved designs are also reconstructed in physical coordinates to check the
observation map, objective scaling, SINR, and power.

Posterior sampling is then checked by holding each trained design fixed and
recertifying it on independent LHS draws using fresh dual variables.
Table~\ref{tab:sampling_adequacy} reports the first independent evaluations
for separated, overlap, and heterogeneous three-target settings.
With $120$ independent Monte Carlo (MC) training draws, the evaluation
certificates reach $\SampMCMax\%$ as the fixed angular spread covers more
beamwidths at larger arrays. Under the LHS sample-size rule, the medians are
$\SampLHSMedianMin$--$\SampLHSMedianMax\%$ and the maximum is $\SampLHSMax\%$.
These comparisons assess the sampling scheme together with its sample count.
Independent repeats and larger evaluation sets change the BCRB by at most
$\SampStabBCRB\%$, with a maximum certificate of $\SampLHSAllMax\%$
across all evaluations of these LHS designs.

\begin{table}[!t]
\centering
\caption{Posterior-sampling validation.}
\label{tab:sampling_adequacy}
{\footnotesize\setlength{\tabcolsep}{2.5pt}
\begin{tabular*}{\columnwidth}{@{\extracolsep{\fill}}lrrccr@{}}\toprule
Scene & $N_t$ & $S$ & MC, $S{=}120$ & LHS & $n$\\\midrule
\multirow{4}{*}{Separated} & 64 & 120 & 3.08 [5.79] & 0.32 [1.36] & 5\\
 & 128 & 120 & 7.26 [8.85] & 0.54 [1.19] & 5\\
 & 256 & 240 & 14.84 [16.60] & 0.34 [0.92] & 5\\
 & 512 & 480 & 20.26 [28.81] & 0.45 [0.56] & 5\\
\midrule
\multirow{4}{*}{Overlap} & 64 & 120 & 4.07 [4.82] & 0.12 [0.14] & 5\\
 & 128 & 120 & 7.75 [20.96] & 0.33 [0.95] & 5\\
 & 256 & 240 & 16.76 [19.99] & 0.46 [0.82] & 5\\
 & 512 & 480 & 26.81 [81.04] & 0.74 [1.36] & 5\\
\midrule
\multirow{2}{*}{$M=3$ (het.)} & 128 & 240 & 9.17 [11.95] & 1.04 [1.28] & 5\\
 & 256 & 480 & 14.59 [21.16] & 0.70 [1.35] & 5\\
\bottomrule\end{tabular*}}
\par\smallskip
\begin{minipage}{\columnwidth}\footnotesize
Median [maximum] gap bounds (\%) are normalized by $f_{\rm up}$ on the first independent evaluation set.
MC and LHS denote Monte Carlo and Latin hypercube sampling.
$S$ counts LHS training draws and $n$ scenes.
Evaluation uses $1000$ draws for $M=3$, $N_t=256$, and $1500$ otherwise.
\end{minipage}
\end{table}

Supplementary Fig.~\ref{fig:exp_certificate_trace} shows how the extension term
triggers expansion in an angular-overlap scene while the reduced gap stays small.

\subsection{Fixed Structures versus Certified Expansion}
\label{sec:exp_structures}
To assess fixed supports under posterior uncertainty,
Table~\ref{tab:obs_comparison} tests the deterministic optimal-beam structure
$\operatorname{orth}[\bm H,\bm A(\bar{\bm\theta}),\dot{\bm A}(\bar{\bm\theta})]$
evaluated at the posterior means~\cite{Fang2025ISACOBS}.
Its nominal dimension is $K+2M$. Each row contains eight paired scenes.
\modelchange{In the separated scenes, its median objective excess over the certified
full-space lower bound is $\ReceiveObsSmallExcess\%$ at $64$ antennas and
$\ReceiveObsLargeExcess\%$ at $512$.}
Across all $\ReceiveObsCount$ nominal-structure tests, including the additional
geometries and task settings, none meets the $1\%$ target.
\modelchange{The corresponding excess for Dynamic Fisher is at most
$\ReceiveObsDynamicMax\%$ in Table~\ref{tab:obs_comparison}.}
Random-geometry results are given in Supplementary Table~\ref{tab:supp_nominal_random}.
Expansion from the nominal structure also meets the certificate target.
For $M=2$, it uses $q=\ReceiveObsTwoOBSQ$, compared with
$\ReceiveObsTwoFisherQ$ from the Fisher basis.

\begin{table}[!t]
\centering
\caption{Nominal and certified designs.}
\label{tab:obs_comparison}
{\footnotesize\setlength{\tabcolsep}{2pt}\renewcommand{\arraystretch}{1.18}
\begin{tabular*}{\columnwidth}{@{\extracolsep{\fill}}lcccc@{}}
\toprule
Scene & $N_t$ & Nominal & \multicolumn{2}{c}{Proposed}\\
\cmidrule(lr){3-3}\cmidrule(l){4-5}
 & & Excess (\%) & $q$ & Max. excess (\%)\\
\midrule
\multirow{4}{*}{Separated} & 64 & 7.9 [7.6, 8.4] & 4--6 & 0.27\\
 & 128 & 16.0 [15.2, 17.3] & 4--6 & 0.14\\
 & 256 & 23.5 [22.2, 24.2] & 4--5 & 0.17\\
 & 512 & 29.1 [28.6, 31.2] & 4--5 & 0.09\\
\midrule
\multirow{4}{*}{Overlap} & 64 & 9.3 [8.9, 11.3] & 6 & 0.05\\
 & 128 & 17.1 [15.2, 19.0] & 6 & 0.08\\
 & 256 & 23.1 [20.6, 26.2] & 6--7 & 0.36\\
 & 512 & 30.8 [26.7, 31.5] & 6--7 & 0.31\\
\bottomrule
\end{tabular*}}
\par\smallskip
\begin{minipage}{\columnwidth}\footnotesize
Excess is normalized by the certified full-space lower bound.
Nominal entries are median [minimum, maximum]. Proposed entries give the
range of $q$ and maximum excess over eight paired scenes.
\end{minipage}
\end{table}

\begin{figure}[!t]
 \centering
 \includegraphics[width=\columnwidth]{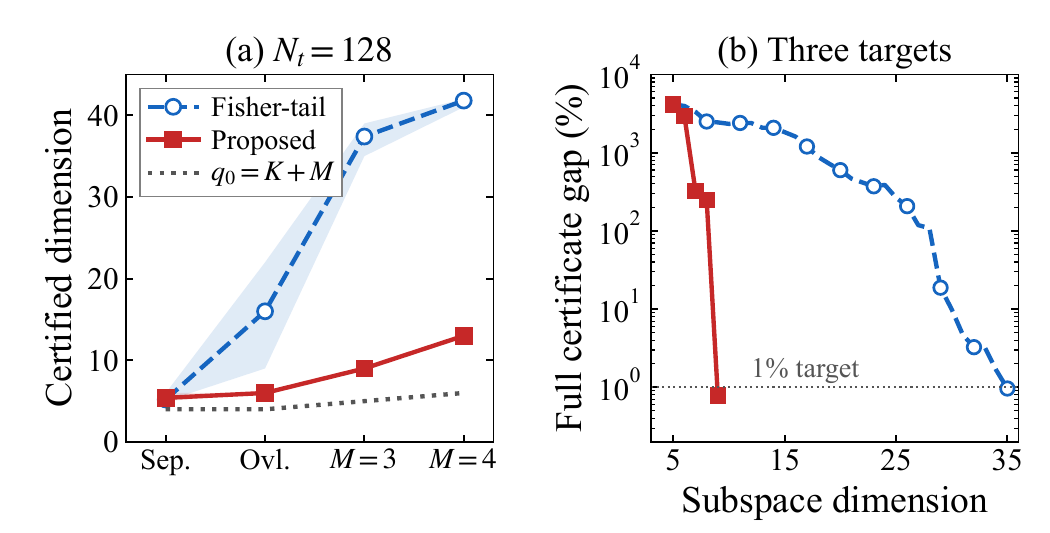}
 \caption{Dynamic and fixed-order expansion. (a) Certified dimension. (b) Three-target path.}
 \label{fig:exp_expansion_ablation}
\end{figure}

The role of the expansion direction is isolated in
Fig.~\ref{fig:exp_expansion_ablation}(a), using five scenes per regime at $N_t=128$.
Both methods compute the same Fisher ordering, start from the same $K+M$
columns, and use identical solves and certificates. Only the rule for appending
directions is varied, with mean final dimensions and min--max bands reported.
The additional $M=3,4$ cases use heterogeneous widths and gain means
specified in the supplement. At $N_t=128$ and a $1\%$ target, three targets
require $q=\StrongMThreeDynamicQ$ dynamically versus $\StrongMThreeTailQ$ with
the fixed ordering (median paired time ratio $\StrongMThreeRatio$), and four
targets require $\StrongMFourDynamicQ$ versus $\StrongMFourTailQ$
($\StrongMFourRatio$). The two methods perform similarly in separated
two-target scenes. With angular overlap, dynamic expansion uses six dimensions,
whereas Fisher-tail requires $9$--$22$.
Across the broader study at $N_t=64,128,256$ and the $1\%$ and $0.1\%$ targets,
all \StrongMechanismCount{} dynamic-expansion runs attain their prescribed
certificate targets. Fisher-tail certifies
\StrongTailSuccess{} cases. The remaining three-target cases at $N_t=256$ and
four-target cases at the $0.1\%$ target reach the time budget. Four-target cases are limited to $N_t\le128$
by evaluation memory.

The trajectory of the first predefined three-target scene is shown in
Fig.~\ref{fig:exp_expansion_ablation}(b). The $1\%$ target is reached at
$q=9$ by dynamic expansion and at $q=35$ by the fixed ordering.
The recorded objective bounds are also used to check strict improvement.
Among the \StrongDynamicExpanded{} runs that expand, all \StrongStrictSteps{}
expansions satisfy $f_{{\rm up},+}<L_{\bm U}$.
Here $L_{\bm U}$ is the old restricted lower bound and $f_{{\rm up},+}$ is
the new feasible objective upper bound. Thus strict improvement is verified
at the finite inner tolerances used in these tests.

\begin{figure}[!t]
 \centering
 \includegraphics[width=0.8\columnwidth]{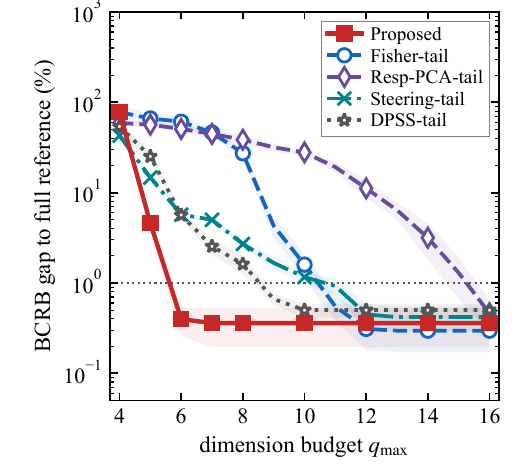}
 \caption{BCRB gaps normalized by the full-space reference.}
 \label{fig:exp_dynamic_ablation}
\end{figure}

At common dimension budgets, Fig.~\ref{fig:exp_dynamic_ablation} compares
Dynamic Fisher with Fisher-tail, Resp-PCA-tail, Steering-tail, and DPSS-tail
on ten heterogeneous $N_t=24$ scenes.
The same MOSEK tolerance policy is used throughout, and both Fisher
variants start from the same four-dimensional basis. For each dimension
budget, the latest available design is used. Certified terminal designs are
retained at larger budgets. Mean gaps and $95\%$ confidence intervals are
therefore computed over all ten scenes, normalized by the full-space
reference objective. At a budget of six, the mean BCRB gap is
$\MultiBudgetDynamicSixGap\%$ for Dynamic Fisher and
$\MultiBudgetDpssSixGap\%$ for DPSS-tail, the best fixed ordering at that
budget.
The dotted line marks the $1\%$ target. Additional results and settings are
given in Supplementary Section~\ref{app:controlled_expansion}.

\subsection{Computation at a Common Certificate}\label{sec:exp_fisher_vs_pca}
\begin{figure}[!t]
 \centering
 \includegraphics[width=0.8\columnwidth]{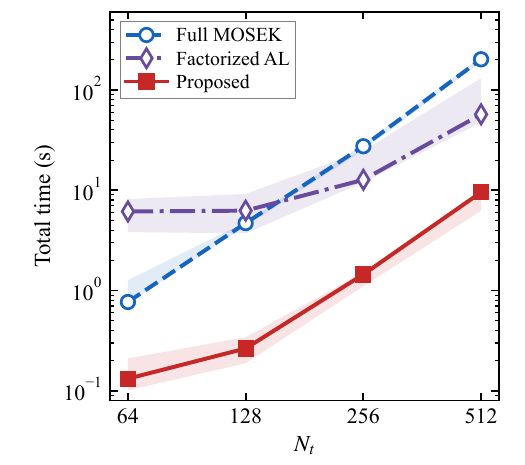}
 \caption{Runtime comparison at a $1\%$ certificate.}
 \label{fig:exp_solver_comparison}
\end{figure}

Fig.~\ref{fig:exp_solver_comparison} compares the total runtime required to
reach a $1\%$ full-space certificate over \RevPilotScenes{} paired scenes at
$N_t=64$--$512$. Medians are shown by curves and min--max ranges by shading. At $512$
antennas, median times are $\RevPilotLargeFullTime$ s for full-space MOSEK,
$\RevPilotLargeBMTime$ s for factorized AL, and $\RevPilotLargeDynamicTime$ s
for Dynamic Fisher. The median scene-wise ratios to Dynamic Fisher are
$\RevPilotLargeFullRatio$ and $\RevPilotLargeBMRatio$, respectively.

Supplementary Table~\ref{tab:supp_initialization} compares initializers under
the same expansion rule and stopping test. In the supplementary probing-rank
study, all tested ranks reach the $1\%$ certificate. Rank two gives the lowest
median paired time ratio to Dynamic Fisher at both array sizes.

The authors' public dual spectral bundle method (SBMD)~\cite{LiaoDingZheng2026}
is compared in Table~\ref{tab:public_specbm} over \StrongPublicScenes{} paired
scenes per setting. Separated and overlap scenes both use
$N_t=64,128,256$, and three-target scenes use $N_t=128,256$.
Parameters are fixed on independent pilot scenes (supplement), and each SBMD
run is given a $120$-s budget starting before model construction and conversion.
The proposed method certifies \StrongPublicDynamicSuccess{} of
\StrongPublicCount{} instances, and SBMD certifies \StrongPublicSuccess{}.
Times include model construction, conversion, restoration, and verification,
and are reported as medians [interquartile range (IQR) widths] over certified runs.
Over jointly certified pairs, the median runtime ratios range from
\StrongPublicRatioMin{} to \StrongPublicRatioMax{}.
The upper value is obtained from three jointly certified pairs at $M=3$, $N_t=128$.
For three targets, the median SBMD terminal gaps at $N_t=128,256$ are
$\StrongPublicHardSmallGap\%$ and $\StrongPublicHardLargeGap\%$, respectively.

\begin{table}[!t]
\centering
\caption{Public SBMD comparison. Times: median [IQR width].}
\label{tab:public_specbm}
{\footnotesize\setlength{\tabcolsep}{1.5pt}\renewcommand{\arraystretch}{1.18}
\begin{tabular*}{\columnwidth}{@{\extracolsep{\fill}}lcrrrr@{}}\toprule
Scene & $N_t$ & Proposed (s) & SBMD (s) & SBMD & Ratio\\
 & & & & success & \\\midrule
\multirow{3}{*}{Separated} & 64 & 0.158 [0.043] & 2.008 [0.805] & 8/8 & 14.7\\
 & 128 & 0.271 [0.050] & 5.993 [0.716] & 8/8 & 24.3\\
 & 256 & 1.448 [0.048] & 30.086 [1.200] & 8/8 & 20.8\\
\midrule
\multirow{3}{*}{Overlap} & 64 & 0.176 [0.012] & 2.029 [0.284] & 8/8 & 11.5\\
 & 128 & 0.360 [0.014] & 5.716 [1.070] & 8/8 & 16.5\\
 & 256 & 1.969 [0.063] & 29.753 [0.679] & 8/8 & 14.9\\
\midrule
\multirow{2}{*}{$M=3$ (het.)} & 128 & 3.091 [0.040] & 80.905 [21.794] & 3/8 & 26.3\\
 & 256 & 19.369 [0.525] & -- & 0/8 & --\\
\bottomrule\end{tabular*}}
\par\smallskip
\begin{minipage}{\columnwidth}\footnotesize
Ratio is the median paired SBMD/Proposed time over jointly certified runs.
Dashes indicate that no SBMD run was certified.
\end{minipage}
\end{table}

Supplementary checks use random geometries and multithreaded full-space solves.
Optimization and certification are timed after shared model construction.
At $512$ antennas, eight-thread full-space MOSEK takes
$\ReceiveThreadsLargeEightRatio$ times the time of single-thread Dynamic
Fisher on three paired scenes.

\subsection{Task Complexity and Array Scaling}\label{sec:exp_array_scaling}
\begin{figure}[!t]
 \centering
 \includegraphics[width=\columnwidth]{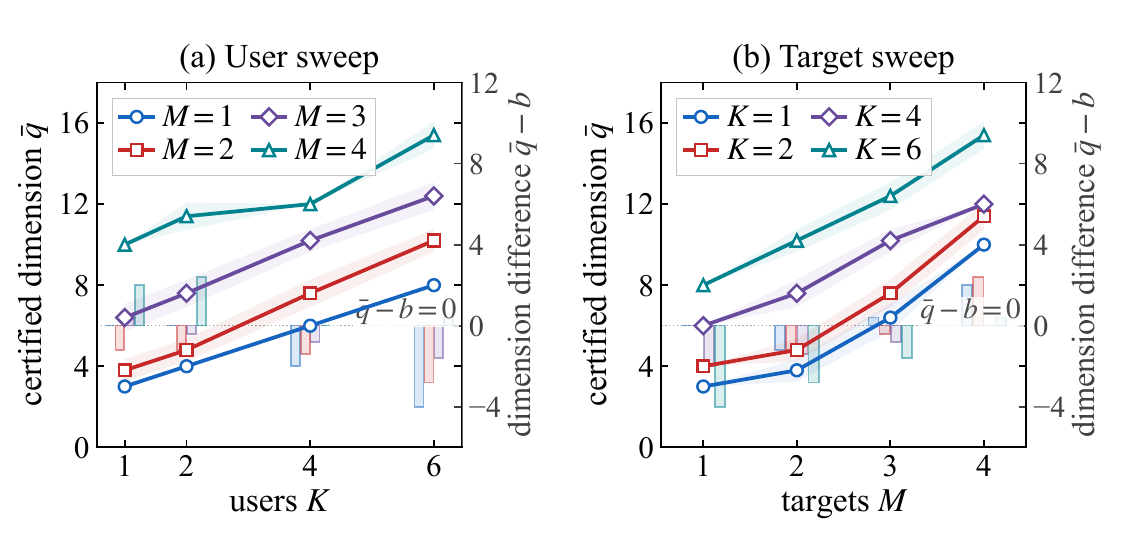}
 \caption{Certified dimension versus task size. Curves show $\bar q$ (left axis),
 and translucent bars show $\bar q-b$ (right axis).}
 \label{fig:exp_task_complexity}
\end{figure}

Fig.~\ref{fig:exp_task_complexity} shows the certified dimension as the user
and target counts are varied at $N_t=64$.
In panel (a), $K\in\{1,2,4,6\}$ is varied
for $M=1,2,3,4$. In panel (b), $M\in\{1,2,3,4\}$ is varied
for $K=1,2,4,6$. Means and $95\%$ confidence intervals are shown over
five scenes per setting. All $80$ designs meet the $1\%$ certificate.
Target centers are equally spaced in $[-35^\circ,35^\circ]$ for $M>1$,
so their separation decreases as $M$ grows.
The mean dimension increases with either task count over this grid.
For example, at $K=6$, it grows from $8.0$ to $15.4$ as $M$ increases
from $1$ to $4$.
The existence bound $b=K+\lfloor\sqrt{K^2+(7M^2+M)/2}\rfloor$ uses the structural
upper bound on $d_J$. \modelchange{By
\textbf{Proposition~\ref{prop:optimal_support}}, it bounds the dimension of a
subspace containing the channel span and an exact optimal design. The generated
subspace may exceed this existence bound.}
For $K=2$ and $M=4$, the mean certified dimension is $11.4$, above the
existence bound of nine.

\begin{figure}[!t]
 \centering
 \includegraphics[width=\columnwidth]{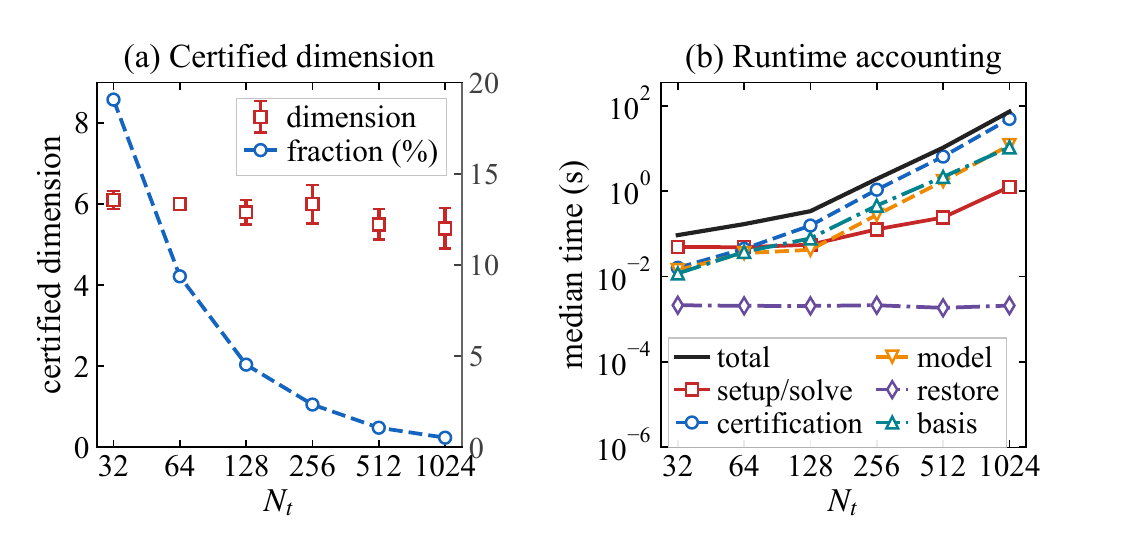}
 \caption{Dimension and runtime versus array size.}
 \label{fig:exp_array_scaling}
\end{figure}

Array scaling is then evaluated with $K=M=2$ and ten scenes shared across
$N_t\in\{32,\ldots,1024\}$ in Fig.~\ref{fig:exp_array_scaling}.
The sampled-SDP optimization target is tightened to $0.1\%$, while
posterior sampling is assessed separately in Table~\ref{tab:sampling_adequacy}
and Supplementary Section~\ref{sec:exp_sampling}.
Panel (a) shows the mean certified dimension with $95\%$ confidence
intervals and the mean fraction of antenna space. Panel (b) reports median
runtime components. \modelchange{The largest certified dimension is $\RevScalingQMax$.
This exceeds the existence bound of six, including the channel span, for
$K=M=2$.} The mean subspace fraction at $1024$ antennas is
$\RevScalingLargeFraction\%$. Median total time increases from
$\RevScalingSmallTime$ s to $\RevScalingLargeTime$ s. At large arrays,
full-dimensional spectral verification accounts for most of this time.

Sensitivity to sensing-information strength is also checked at $256$
antennas by varying $\alpha_s\in\{10^2,10^3,10^4\}$. All fifteen dynamic
designs attain the $1\%$ certificate with $q=4$--$5$, and all nine full-space
controls also certify.

\section{Discussion and Scope}\label{sec:discussion}
The certificate bounds the objective gap for the specified sampled information
operator and feasible set. Posterior integration and model mismatch are
separate error sources. The independent-sample checks in
Section~\ref{sec:exp_validation} assess stability on a second sampled problem.
For model mismatch, assuming
widths of $3^\circ$--$12^\circ$ for a $6^\circ$ posterior changes the held-out
BCRB by at most $\ReceiveMismatchMax\%$ on average (supplement). A population
statement remains conditional on a uniform information error bound.

Under the conditions of \textbf{Theorem~\ref{thm:strict_expansion}}, exact solves
yield strict improvement after certificate-driven expansion.
At finite inner tolerances, improvement is checked
using $f_{{\rm up},+}<L_{\bm U}$, as in the controlled expansion experiments.

Full-space spectral verification remains the main cost at large arrays.
Faster certified spectral enclosures could reduce this cost for both subspace
and factorized implementations.

\section{Conclusion}\label{sec:conclusion}
A certified dynamic subspace method was developed for Bayesian ISAC transmit
covariance design. Starting from a Fisher-sensitive basis, restricted
problems are solved and the basis is expanded until the full-space gap bound
meets the requested accuracy.
\methodchange{Under a strict communication-power margin, joint dual uniqueness
is established from the SINR block structure. Consequently, certificate-driven
expansion strictly improves the optimal BCRB after an exact restricted solve.}
Primal restoration and lossless user-beam recovery retain the full-space certificate for
the returned feasible transmit design.
\methodchange{Controlled multi-target experiments show that adaptive directions can reduce
the dimension required by a fixed Fisher ordering. At $512$ antennas and a
$1\%$ certificate target, the
median paired speedup over full-space MOSEK is $\RevPilotLargeFullRatio$,
including model construction and full-space verification.
In the array-scaling study with $K=M=2$, all tested arrays from $32$ to
$1024$ antennas meet the sampled-SDP $0.1\%$ certificate with $q\le\RevScalingQMax$.}

\IEEEtriggeratref{14}

\IEEEtriggeratref{0}

\clearpage
\raggedbottom
\raggedend                       
\allowdisplaybreaks[1]            
\twocolumn[{%
\begin{center}
{\huge\textbf{Supplementary Material}\par}
\vspace{4pt}
{\normalsize\textit{Certified Dynamic Subspace Optimization for Bayesian ISAC
Design}\par}
\vspace{12pt}
\end{center}
}]
\setcounter{page}{1}
\renewcommand{\thepage}{S\arabic{page}}
\setcounter{section}{0}
\renewcommand{\thesection}{S\arabic{section}}
\renewcommand{\theequation}{S\arabic{equation}}
\setcounter{equation}{0}
\renewcommand{\thetheorem}{S\arabic{theorem}}
\setcounter{figure}{0}
\renewcommand{\thefigure}{S\arabic{figure}}
\setcounter{table}{0}
\renewcommand{\thetable}{S\arabic{table}}

\newcommand{\SuppCite}[1]{\cite{#1}}

\makeatletter
\setlength{\@fptop}{0pt}
\setlength{\@fpsep}{8pt}
\setlength{\@fpbot}{0pt plus 1fil}
\setlength{\@dblfptop}{0pt}
\setlength{\@dblfpsep}{8pt}
\setlength{\@dblfpbot}{0pt plus 1fil}
\makeatother

\providecommand{\modelchange}[1]{{#1}}
\providecommand{\methodchange}[1]{{#1}}

\theoremstyle{definition}
\newtheorem{suppproposition}{Proposition}
\renewcommand{\thesuppproposition}{S\arabic{suppproposition}}

This supplement provides the proofs, numerical mappings, and supporting
comparisons used in the main text. Unprefixed results refer to the main
paper. Supplementary equations and tables use the prefix S.

\section{Subspace Structure and Proofs}
\label{app:subspace_structure}\label{app:proofs}

\subsection{Exact Projection and Posterior Moments}
For fixed target responses, let $\bm P_0$ project onto
$\mathcal S_0=\spanop[\bm H,\bm A_0,\dot{\bm A}_0]$.
Every column of the receive-array factors $\bm\Gamma_r$ belongs to this
space, whose dimension is at most $K+2M$.
\begin{suppproposition}[Exact reduction for fixed responses]
\label{prop:posterior_response_observability}
Replacing each feasible block by $\bm Q_i^0=\bm P_0\bm Q_i\bm P_0$
preserves the fixed-response BIM and all SINR margins. Total power is not
increased. Hence the fixed-response problem admits an optimum on
$\mathcal S_0$.
\end{suppproposition}
\begin{IEEEproof}
Congruence preserves PSD. Since $\bm P_0\bm h_k=\bm h_k$ and
$\bm P_0\bm\Gamma_r=\bm\Gamma_r$, both
$\bm h_k^H\bm Q_i^0\bm h_k=\bm h_k^H\bm Q_i\bm h_k$ and
$\bm\Gamma_r^H\bm Q^0\bm\Gamma_r=\bm\Gamma_r^H\bm Q\bm\Gamma_r$ hold.
Moreover, $\tr(\bm Q_i^0)=\tr(\bm P_0\bm Q_i)\le\tr(\bm Q_i)$.
The epigraph block can therefore be retained with the same objective.
\end{IEEEproof}

For posterior averaging, write $\bm g_{a,r}$ for column $a$ of
$\bm\Gamma_r$ and define
\begin{equation}\label{eq:exact_task_span}
\begin{aligned}
 \bm R_{ab}&=\sum_r\E[\bm g_{b,r}\bm g_{a,r}^H\mid z],\\
 \mathcal S_{\rm task}&=\range(\bm H)+
 \sum_{a,b}\bigl(\range(\bm R_{ab})+\range(\bm R_{ab}^H)\bigr).
\end{aligned}
\end{equation}
Projection onto this space gives $\bm P\bm R_{ab}\bm P=\bm R_{ab}$.
All posterior response traces are preserved by the preceding argument.
This construction is useful for reduction when its dimension is below $N_t$.

\subsection{Proof of Proposition~\ref{prop:posterior_moment_full_rank}}
For $\bm x\ne0$, the quadratic form $\bm x^H\bm R_a\bm x$ is the
posterior integral of $|\bm x^H\bm a(u)|^2$. If it vanishes, positivity
of the density on the stated interval forces the polynomial
$\bm x^H\bm a(u)$ to vanish almost everywhere there. A polynomial of
degree at most $N_t-1$ with infinitely many distinct nodes
$e^{\mathrm j\pi u}$ is identically zero. This contradicts $\bm x\ne0$,
so $\bm R_a\succ0$. Preservation for every rank-one PSD covariance gives
$\bm R_a=\bm P\bm R_a\bm P$. Any nonzero vector in $\ker\bm P$ would
then have zero quadratic form under $\bm R_a$. Thus $\bm P=\bm I$.

\subsection{Feasibility and Nested Supports}
For \textbf{Lemma~\ref{lem:base_feasibility}}, projection onto a space containing
$\range(\bm H)$ preserves every channel quadratic form and does not
increase trace, as above. The projected BIM is positive definite under
the stated assumption. Choosing its angle inverse block as $\bm Z$
satisfies the Schur constraint and proves reduced feasibility.
For \textbf{Lemma~\ref{lem:feasible_expansion}}, let
$\bar{\bm Q}_{i,+}=\bm U_+^H\bm Q_i\bm U_+$. Nestedness gives
$\bm U_+\bar{\bm Q}_{i,+}\bm U_+^H=\bm Q_i$.
The physical design, constraints, and objective are unchanged.

\subsection{Proof of Theorem~\ref{thm:dynamic_certificate}}
Write the admissible dual candidate without hats. For any feasible blocks
$\bm S_i$ and Schur matrix $\bm\Phi(\bm S,\bm Z)$, the Lagrangian identity is
\begin{equation}\label{eq:supp_lagrangian_identity}
\begin{aligned}
 &\tr(\bm Z)-d_0-\sum_i\tr(\bm G_i\bm S_i)\\
 &\quad=\tr(\bm Y\bm\Phi)+\tr((\bm I-\bm Y_Z)\bm Z)\\
 &\qquad+\sum_k\mu_k c_k+
 \nu\bigl(P-\sum_i\tr(\bm S_i)\bigr)\ge0.
\end{aligned}
\end{equation}
Each term is nonnegative by primal feasibility and dual admissibility.
Since $\bm G_i\succeq-\delta_{\rm full}\bm I$ and the total trace is
at most $P$, taking the infimum gives
$p^\star\ge d_0-P\delta_{\rm full}$.
For restricted blocks $\bm S_i=\bm U\bar{\bm S}_i\bm U^H$, the same
argument uses $\bm U^H\bm G_i\bm U\succeq-\eta_{\bm U}\bm I$ and gives
$p_{\bm U}^\star\ge d_0-P\eta_{\bm U}$.
Subtracting these bounds from the feasible incumbent objective proves
both gap bounds and
\begin{equation}\label{eq:supp_gap_decomposition}
 \Delta_{\rm full}=\Delta_{\rm red}
                  +P(\delta_{\rm full}-\eta_{\bm U}).
\end{equation}
The extension term is nonnegative by restriction of the Rayleigh quotient.

\subsection{Outside Directions and Finite Termination}
For \textbf{Lemma~\ref{lem:outside_direction}}, if a unit minimum eigenvector
$\bm v^\star$ belonged to $\range(\bm U)$, then
\begin{equation}\label{eq:supp_outside_rayleigh}
 -\delta_{\rm full}
 =(\bm v^\star)^H\bm G_{i^\star}\bm v^\star
 \ge-\eta_{\bm U},
\end{equation}
contradicting $\delta_{\rm full}>\eta_{\bm U}$.
Its nonzero orthogonal residual therefore defines $\bm d^\star$.
The enlarged space contains $\bm v^\star$, so its restricted slack has
minimum eigenvalue at most $-\delta_{\rm full}$.

For \textbf{Corollary~\ref{cor:finite_termination}}, an expansion is triggered only
when $\widehat\Delta_{\rm red}\le\varepsilon_{\rm in}
<\varepsilon_{\rm out}<\widehat\Delta_{\rm full}$.
The gap decomposition implies $\delta_{\rm full}>\eta_{\bm U}$.
Each expansion thus adds a dimension. After at most $N_t-q_0$ expansions,
the two defects coincide. The assumed reduced accuracy then implies the
full stopping test, and \textbf{Theorem~\ref{thm:dynamic_certificate}} applies.

\subsection{\methodchange{Proof of Theorem~\ref{thm:strict_expansion}(i)}}
\begin{IEEEproof}
\methodchange{We first establish attainment and strong duality. Since
$P_{\rm c}(\bm U)<P$, a communication-feasible collection with power below
$P$ exists. Scaling it by a factor slightly larger than one makes every
SINR margin strict because $\gamma_k\sigma_k^2>0$. A sufficiently small
positive identity can then be added to each reduced covariance while
retaining strict SINR and power margins. The BIM is positive definite by
\textbf{Assumption~\ref{ass:information_regular}}. Choosing
$\bm Z=\bm E_\theta\bm J_{\rm B}^{-1}\bm E_\theta^T+\epsilon\bm I_M$
with $\epsilon>0$ makes the Schur constraint strict. Slater's condition
therefore holds. The primal optimum is attained by compactness of the
covariance feasible set and continuity of the objective. The dual optimum
is attained with the same value.

Fix any restricted primal optimum. All covariance and channel variables
in this proof are expressed in reduced coordinates, with bars omitted.
Write $\bm C_{k,i}^{\bm U}=\bm U^H\bm C_{k,i}\bm U$.
Let $\bm J$ denote the optimal BIM. The tight epigraph satisfies
$\bm Z=\bm E_\theta\bm J^{-1}\bm E_\theta^T\succ0$.
Every optimal dual triple is complementary to this fixed primal optimum.
Complementarity with $\bm Z$ gives $\bm Y_Z=\bm I_M$.
Complementarity with the Schur matrix then fixes
\begin{equation}\label{eq:supp_unique_schur_dual}
 \bm Y=
 \begin{bmatrix}\bm J^{-1}\bm E_\theta^T\\-\bm I_M\end{bmatrix}
 \begin{bmatrix}\bm E_\theta\bm J^{-1}&-\bm I_M\end{bmatrix}.
\end{equation}
Thus $\bm Y$ is identical for every optimal dual triple.

Consider two such triples and write
$a=\nu^{(1)}-\nu^{(2)}$ and
$\bm b=\bm\mu^{(1)}-\bm\mu^{(2)}$.
Both reduced covariance slacks annihilate the same optimal blocks, so
\begin{equation}\label{eq:supp_dual_difference}
 \left(a\bm I_q-\sum_k b_k\bm C_{k,i}^{\bm U}\right)\bm Q_i=0,
 \qquad i\in\mathcal I.
\end{equation}
For each user block, let $p_i=\tr(\bm Q_i)>0$. Define
$\bm A_{\rm c}\in\mathbb R^{K\times K}$ by
\begin{equation}\label{eq:supp_sinr_matrix}
 [\bm A_{\rm c}]_{ki}
 =\tr(\bm C_{k,i}^{\bm U}\bm Q_i)
 =\begin{cases}
  \bm h_k^H\bm Q_k\bm h_k,&i=k,\\
  -\gamma_k\bm h_k^H\bm Q_i\bm h_k,&i\ne k.
 \end{cases}
\end{equation}
SINR feasibility and positive noise give
\begin{equation}\label{eq:supp_sinr_diagonal_dominance}
 [\bm A_{\rm c}]_{kk}-\sum_{i\ne k}|[\bm A_{\rm c}]_{ki}|
 \ge\gamma_k(\bm h_k^H\bm Q_s\bm h_k+\sigma_k^2)>0.
\end{equation}
Hence $\bm A_{\rm c}$ is strictly diagonally dominant, with positive
diagonal and nonpositive off-diagonal entries. Its inverse is nonnegative
with positive diagonal. To see this, write
$\bm A_{\rm c}=\bm D(\bm I-\bm N)$ with diagonal $\bm D\succ0$,
$\bm N\ge0$, and $\|\bm N\|_\infty<1$. The Neumann series gives
$\bm A_{\rm c}^{-1}=\sum_{j\ge0}\bm N^j\bm D^{-1}$.
Taking traces of the user-block identities yields
\begin{equation}\label{eq:supp_multiplier_linear_system}
 \bm A_{\rm c}^T\bm b=a\bm p.
\end{equation}
If $a=0$, then $\bm b=0$, and the two triples are identical.

Suppose $a\ne0$ and set
\begin{equation}\label{eq:supp_normalized_dual_difference}
 \bm\lambda=\bm b/a=\bm A_{\rm c}^{-T}\bm p>0,
 \qquad
 \bm D_i=\bm I_q-\sum_k\lambda_k\bm C_{k,i}^{\bm U}.
\end{equation}
Then $\bm D_i\bm Q_i=0$ for every block. For the probing block,
\begin{equation}\label{eq:supp_probing_difference_slack}
 \bm D_s=\bm I_q+\sum_k\lambda_k\gamma_k\bm h_k\bm h_k^H\succ0,
\end{equation}
so $\bm Q_s=0$. Each user block satisfies
\begin{equation}\label{eq:supp_user_difference_slack}
 \bm D_k=\bm D_s-(1+\gamma_k)\lambda_k\bm h_k\bm h_k^H.
\end{equation}
Since $\bm Q_k\ne0$, this matrix is singular. Congruence by
$\bm D_s^{-1/2}$ gives
$\bm I_q-(1+\gamma_k)\lambda_k\bm z_k\bm z_k^H$, where
$\bm z_k=\bm D_s^{-1/2}\bm h_k$. Singularity forces its nonunit
eigenvalue to equal zero. Thus $\bm D_k\succeq0$ for every user.

The vector $\bm\lambda>0$ and slacks $\bm D_i\succeq0$ are dual feasible
for the minimum communication-power problem defining $P_{\rm c}(\bm U)$.
Subtracting the two sets of scalar complementary-slackness identities and
dividing by $a\ne0$ gives
\begin{equation}\label{eq:supp_difference_complementarity}
 \lambda_k c_k^{\bm U}(\{\bm Q_i\})=0,
 \qquad \sum_i\tr(\bm Q_i)=P.
\end{equation}
Taking traces of $\bm D_i\bm Q_i=0$ and summing over the blocks yields
\begin{equation}\label{eq:supp_communication_power_equality}
 P=\sum_i\tr(\bm Q_i)=\sum_k\lambda_k\gamma_k\sigma_k^2.
\end{equation}
The fixed primal collection and $\bm\lambda$ therefore certify
$P_{\rm c}(\bm U)=P$, contradicting the strict power margin. Hence $a=0$
and $\bm b=0$. The optimal dual triple is unique.}
\end{IEEEproof}

\subsection{\methodchange{Proof of Theorem~\ref{thm:strict_expansion}(ii)}}
\begin{IEEEproof}
\methodchange{The repair $\widetilde\nu=\nu+\eta_{\bm U}$ gives a feasible restricted
dual triple. Its gap to the feasible restricted primal point is
$\Delta_{\rm red}=0$, so both points are optimal. By
\textbf{Theorem~\ref{thm:strict_expansion}(i)},
$(\bm Y,\bm\mu,\widetilde\nu)$ is the unique restricted dual optimum.

Its full-space covariance slacks are
$\widetilde{\bm G}_i=\bm G_i+\eta_{\bm U}\bm I$.
By \textbf{Lemma~\ref{lem:outside_direction}},
$\bm v^\star\in\range(\bm U_+)$, and
\begin{equation}\label{eq:supp_expanded_dual_violation}
 (\bm v^\star)^H\widetilde{\bm G}_{i^\star}\bm v^\star
 =-\delta_{\rm full}+\eta_{\bm U}<0.
\end{equation}
Thus this triple is not dual feasible in the enlarged space.

The strict communication-power margin is retained under expansion.
The Slater construction in the preceding proof therefore applies to
$\bm U_+$, giving an attained dual optimum. Nested primal feasibility gives
$p_{\bm U_+}^\star\le p_{\bm U}^\star$. Suppose equality holds.
Restricting an enlarged optimal dual triple to $\range(\bm U)$ retains
dual feasibility and its objective value, so it is optimal for the old
problem. Uniqueness forces it to equal
$(\bm Y,\bm\mu,\widetilde\nu)$, contradicting the negative quadratic
form above. Equality is impossible, proving
$p_{\bm U_+}^\star<p_{\bm U}^\star$.}
\end{IEEEproof}

\methodchange{With finite inner tolerances, strict improvement can also be checked
from the computed bounds. Let $L_{\bm U}$ be a valid lower bound on the
old restricted optimum and let $f_{{\rm up},+}$ be the objective upper bound
of a feasible design in the enlarged space. Whenever
$f_{{\rm up},+}<L_{\bm U}$,
\begin{equation}\label{eq:supp_finite_precision_improvement}
 p_{\bm U}^\star-p_{\bm U_+}^\star
 \ge L_{\bm U}-f_{{\rm up},+}>0.
\end{equation}
This is an optional a posteriori check using the same objective bounds.}

\subsection{Proof of Corollary~\ref{cor:rank_one_transfer}}
Positive noise and SINR feasibility give
$a_k=\bm h_k^H\bm Q_k\bm h_k\ge\gamma_k\sigma_k^2>0$.
The Schur complement, or Cauchy--Schwarz applied through $\bm Q_k^{1/2}$,
gives
\begin{equation}\label{eq:supp_rank_one_psd_order}
 \bm0\preceq\widetilde{\bm Q}_k
 =\bm Q_k\bm h_k\bm h_k^H\bm Q_k/a_k\preceq\bm Q_k.
\end{equation}
Moving the PSD differences to the probing block retains the total
covariance. Also $\bm h_k^H\widetilde{\bm Q}_k\bm h_k=a_k$, so all
SINR margins, the BIM, and power are unchanged. All blocks remain in the
original support. The same epigraph and admissible dual candidate give the
same objective bounds and certificate.

\subsection{Fisher Sensitivity}
A Hermitian-moment representation of the information map is
\begin{equation}\label{eq:bim_linear_factorization}
 \bm J_{\rm B}=\bm J_{\rm prior}
 +\sum_\ell\bm F_\ell\tr(\widehat{\bm R}_\ell\bm Q),
\end{equation}
where $\bm F_\ell$ are real symmetric and $\widehat{\bm R}_\ell$ are
Hermitian. Cross moments need not be PSD. Differentiation gives
$\bm T_{\rm F}=\sum_\ell\tr(\bm F_\ell\bm S_{\rm B})
\widehat{\bm R}_\ell=\mathcal J_z^*(\bm S_{\rm B})$.
Positivity of the information map implies
$\bm v^H\bm T_{\rm F}\bm v
=\tr(\bm S_{\rm B}\mathcal J_z(\bm v\bm v^H))\ge0$.
Ky Fan's principle\SuppCite{GolubVanLoan2013} therefore gives
\begin{equation}\label{eq:supp_ky_fan_tail}
 \min_{\bm U^H\bm U=\bm I_q}
 \tr((\bm I-\bm U\bm U^H)\bm T_{\rm F})
 =\tr(\bm T_{\rm F})-\sum_{j=1}^q\lambda_j(\bm T_{\rm F}).
\end{equation}
The channel-orthogonal version uses $\bm T_{{\rm F},\perp}$.
At an optimum, Schur complementarity gives
$\bm Y_J=\bm S_{\rm B}$, recovering the sensing-direction operator
of~\SuppCite{AttiahYu2025Duality}.

\section{Numerical Implementation}
\label{app:solver_details}
The objective-gap bounds require feasible covariances and admissible dual
variables in the original coordinates. The transformations and numerical
restoration used to obtain them are described below.

\subsection{Coordinate Maps and Dual Recovery}
\label{app:coordinate_scaling}\label{app:dual_repair}
The same invertible parameter scaling is used for all methods. It keeps the
spectral roundoff bounds meaningful at large $\alpha_s$. Physical objectives,
bounds, and multipliers are recovered exactly, with relative gaps unchanged.
For the Gaussian experimental prior, partition
$\bm J_{\rm prior}=[\begin{smallmatrix}\bm A&\bm B\\
\bm B^T&\bm C\end{smallmatrix}]$ by angle and gain coordinates.
If $\bm A-\bm B\bm C^{-1}\bm B^T=\bm O\bm\Lambda\bm O^T$, set
\begin{equation}\label{eq:receive_coordinate_map}
 \bm T=\begin{bmatrix}c\bm O&\bm0\\
 -c\bm C^{-1}\bm B^T\bm O&\bm C^{-1/2}\end{bmatrix}.
\end{equation}
Define $\bm D_T=\operatorname{blkdiag}(\bm T,c^{-1}\bm O)$.
Let $\bm T_1=\bm T|_{c=1}$ and
$\bm J_{\rm iso}=\bm J_{\rm B}(P\bm I/N_t)$.
The scale is $c=(\max_{j\le M}[\bm T_1^T\bm J_{\rm iso}\bm T_1]_{jj})^{-1/2}$.
Then $\bm J'=\bm T^T\bm J\bm T$ has diagonal prior
$\diag(c^2\bm\Lambda,\bm I)$, and the physical BCRB is $c^2$ times
the angle BCRB in these coordinates. Analytic gain-variance terms are
folded into response factors before this congruence.

The solver additionally uses
$\bm T_\varsigma=\operatorname{blkdiag}(\varsigma^{-1}\bm I_M,\bm I_{2M})$.
The scale $\varsigma$ is the larger of one and the median derivative norm
weighted by the root mean-square gain. The conditioned BIM is denoted by
$\bm J_E=\bm T_\varsigma\bm J'\bm T_\varsigma$.
With $\tau_E^2$ the larger of one and the median scaled angular-prior entry,
the Schur constraint is
\begin{equation}\label{eq:supp_scaled_schur_branch}
 \bm\Phi_E=\begin{bmatrix}
 \bm T_\varsigma\bm J'\bm T_\varsigma&\tau_E\bm E_\theta^T\\
 \tau_E\bm E_\theta&\bm Z_E
 \end{bmatrix}\succeq0,\qquad
 f_E=\tau_E^{-2}\tr(\bm Z_E).
\end{equation}
Writing $\bm Z'=c^{-2}\bm O^T\bm Z\bm O$ gives
$\bm Z_E=\tau_E^2\varsigma^2\bm Z'$.
The physical epigraph objective is $sf_E$, where $s=c^2/\varsigma^2$.

The raw Schur dual is projected to a positive numerical floor, giving
$\bm Y_{E,+}$. With $a_E=\tau_E^{-2}$, its admissible repair is
\begin{equation}\label{eq:supp_scaled_dual_repair}
 \kappa_E=\max\{1,\lambda_{\max}((\bm Y_{E,+})_Z)/a_E\},\quad
 \bm Y_E=\bm Y_{E,+}/\kappa_E.
\end{equation}
An upper spectral enclosure and outward rounding are used in the cap.
The SINR multipliers are clipped to $\bm\mu_\varsigma\ge0$.
The raw power multiplier is set to zero and reoptimized analytically as
in \textbf{Remark~\ref{rem:optimal_power_multiplier}}.
Let
\begin{equation}\label{eq:supp_physical_congruence}
 \bm D=\bm D_T\operatorname{blkdiag}(\bm T_\varsigma,
 \varsigma\bm I_M)\operatorname{blkdiag}(\bm I_{3M},\tau_E\bm I_M).
\end{equation}
The physical return map is
\begin{equation}\label{eq:receive_physical_dual_map}
 \bm Y=s\bm D\bm Y_E\bm D^T,\qquad
 (\bm\mu,\nu)=s(\bm\mu_\varsigma,\nu_\varsigma).
\end{equation}
Congruence preserves PSD and
$\bm Y_Z=\tau_E^2\bm O(\bm Y_E)_Z\bm O^T\preceq\bm I$.
The adjoint, slacks, and dual constant obey the same return scaling.
All absolute objective bounds are multiplied by $s$, so relative gaps
are invariant. Transmit covariances and communication constraints are unchanged.

\subsection{Primal Restoration and Arithmetic Bounds}
\label{app:numerical_bounds}
Let $u$ be machine precision and $\gamma_n=nu/(1-nu)$.
Each reduced covariance $\bm X$ is projected to the numerical floor
$8\gamma_{64q+128}q\max\{1,\|\bm X\|_F\}$.
For $\bm H_U=\bm U^H\bm H$, the zero-forcing matrix is
$\bm W=\bm H_U(\bm H_U^H\bm H_U)^{-1}$.
The anchor uses blocks $a\gamma\sigma^2\bm w_k\bm w_k^H$ and zero probing.
If its power at $a=1$ is $P_{\rm ZF}<P$, then
$a=\min\{2,(1+P/P_{\rm ZF})/2\}$ gives strict SINR and power margins.
Mixing by \eqref{eq:primal_mixing} restores all affine constraints.
Restoration fails if the numerical anchor check fails. When zero forcing is
unavailable, another strictly feasible reference can be used.

For a computed decomposition $\bm B\approx\bm V\bm\Lambda\bm V^H$,
let $e_R$ and $e_O$ bound its Frobenius residual and orthogonality defect.
The spectral enclosure is
\begin{equation}\label{eq:supp_spectral_enclosure}
 \lambda_{\max}(\bm B)\le\lambda_{\max}(\bm\Lambda)
 +e_R+\|\bm\Lambda\|_2e_O,
\end{equation}
with the analogous lower bound. Residual norms are inflated by
$\gamma_{32n+64}$ times matrix-product norm bounds.
Slack formation errors are bounded using component norms before cancellation.
These enclosures verify covariance PSD, the Schur-dual cap, and the
analytically repaired power multiplier.

For the supplied information operator, let
$f(\bm Q)=\tr(\bm E_\theta\bm J^{-1}\bm E_\theta^T)$ denote the angle BCRB
in the coordinates of $\bm J$. In physical coordinates, this is the criterion
$\Psi_{\rm B}(\bm Q,z)$ evaluated with the supplied information operator.
If $\bm J\bm T_0\approx\bm E_\theta^T$ has residual bound $r$, including
assembly error, and $\bm J\succeq\mu\bm I$ is verified, then
\begin{equation}\label{eq:supp_primal_upper}
 f(\bm Q)\le\tr(\bm E_\theta\bm T_0)+\sqrt M\,r/\mu=:f_{\rm up}.
\end{equation}
Indeed, the solution error is at most $r/\mu$ in Frobenius norm.
Scalar evaluation roundoff is added outward.
The value of $\mu$ is the larger of the prior lower bound and the
verified lower eigenvalue of $\bm J$, after assembly-error allowance.
The dual bound is decreased conservatively. The bounds are returned to
physical units using the preceding scalings and apply to the supplied
sampled information operator.

Before dynamic augmentation, the selected eigenvector is orthogonalized
twice against the current basis. Expansion fails if the outside residual
norm is below $10^{-9}$. After reoptimization, both gap bounds are
recomputed from the new dual variables. Finite-tolerance strict improvement
is checked separately through $f_{{\rm up},+}<L_{\bm U}$.
The restored reduced covariance blocks and basis are returned and define
the lifted covariances. \textbf{Corollary~\ref{cor:rank_one_transfer}} gives beam recovery
in exact arithmetic. For floating-point beam outputs, the represented
covariances must be rechecked before the certificate is transferred.

For the evaluated designs, the restored covariances satisfy the original
PSD, SINR, and power constraints. Relative objective changes from restoration
do not exceed $0.002\%$, $0.3\%$, and $0.9\%$ for the proposed method,
full-space solves at tolerance $10^{-4}$, and SBMD, respectively.
Certificates are evaluated after restoration.

\subsection{Native MOSEK Mapping}
\label{app:mosek_mapping}
For a Hermitian coefficient $\bm C$, define
\begin{equation}\label{eq:supp_realification_map}
 \mathcal R(\bm C)=\frac12\begin{bmatrix}\Re\bm C&-\Im\bm C\\
 \Im\bm C&\Re\bm C\end{bmatrix}.
\end{equation}
A real PSD variable $\bm X$ represents
\begin{equation}\label{eq:supp_complex_covariance_recovery}
 \bar{\bm Q}=\tfrac12(\bm X_{11}+\bm X_{22})
 +\tfrac{\mathrm j}{2}(\bm X_{21}-\bm X_{12}).
\end{equation}
This map is onto the Hermitian PSD cone and satisfies
$\tr(\mathcal R(\bm C)\bm X)=\Re\tr(\bm C\bar{\bm Q})$.
The Schur epigraph is a separate real PSD variable with affine equality
constraints on its information and off-diagonal blocks.
Its cone dual supplies the raw Schur multiplier. SINR and power
upper-bound duals are negated to obtain the nonnegative multipliers.
Before repair, the explicit dual objective is checked against MOSEK's value
\SuppCite{MosekFusion11}.
Solver tolerances follow Section~\ref{sec:exp_setup} of the main text.
If a mapping or certificate check fails, the next tighter tolerance is used.

\subsection{Factorized AL Baseline}
\label{app:factorized_comparison}
The factorized baseline is evaluated with the same full-space certificate
as the proposed method. Its AL implementation uses rank-one user factors
and probing rank $r_s=2$ by default.
Initialization uses zero-forcing user beams and channel-orthogonal
Fisher directions, with $90\%$ of the remaining power assigned to probing.
The objective is normalized by $f(P\bm I/N_t)$.
SINR and power constraints are written as $c_j\le0$, with residuals normalized
by $\gamma\sigma^2$ and $P$. The augmented objective is
\begin{equation}\label{eq:supp_factorized_al_objective}
 \frac{f(\bm Q)}{f(P\bm I/N_t)}+
 \frac{1}{2\rho}\sum_j\bigl([\lambda_j+\rho c_j]_+^2-\lambda_j^2\bigr).
\end{equation}
Factor subproblems are solved by L-BFGS with analytic gradients and direct
posterior products. Each solve allows 350 iterations with function/gradient
tolerances $10^{-13}$/$10^{-8}$.
After each outer step, $\lambda_j\leftarrow[\lambda_j+\rho c_j]_+$.
The penalty starts at one and increases fivefold, up to $10^7$, when the
maximum violation fails to decrease by a factor $0.3$.
The limits of 30 outer steps and 120 seconds are checked after each step.

At each checkpoint, nonnegative multipliers are fitted to factor
stationarity by least squares. With
$\bm T_0=\bm J_E^{-1}(\tau_E\bm E_\theta^T)$, the Schur dual candidate is
$\tau_E^{-2}[\bm T_0^T,-\bm I]^T[\bm T_0^T,-\bm I]$.
Restoration uses the span of factors and channels, followed by the common
full-space certificate. The final restored result is reported.
The reported support dimension includes the restoration span.

\subsection{Public Spectral-Bundle Baseline}
\label{app:public_specbm}
The authors' public SpecBM implementation\SuppCite{LiaoDingZheng2026} is used
in its dual form (SBMD). The covariance design is represented by an equivalent
real PSD cone of size $2(K+1)N_t+4M+K+1$.
It contains covariance, Schur, and scalar-slack principal blocks.
Off-block entries are free and absent from all affine terms.
Principal-block extraction and block-diagonal embedding show equivalence
to the original product cone.

Covariances are divided by $2P$. A strict zero-forcing/Fisher anchor
defines the diagonal Schur congruence
$\bm D_A=\operatorname{blkdiag}(\bm D_J,\bm D_Z)$ with unit anchor diagonal.
The objective is divided by $f_A=\tau_E^{-2}\tr(\bm Z_{E,A})$.
Scalar slacks are $c_k/(P\|\bm h_k\|_2^2)$ and $1-\tr(\bm Q)/P$,
with $c_k$ defined in \eqref{eq:posterior_sinr_affine}.
Equality rows are scaled by inverse Frobenius norms. For SINR and power
rows, the scale is $1/\max\{|b_j|,\|\bm A_j\|_F/100\}$.
The penalty is $1.1T_A$, with $T_A$ inflated by $1+10^{-8}$ and
\begin{equation}\label{eq:supp_specbm_trace_bound}
\begin{aligned}
 T_A={}&K+2+\tr(\bm D_J^2\bm J_{{\rm prior},c})\\
 &+P\lambda_{\max}(\mathcal J_c^*(\bm D_J^2))
 +\tau_E^2 f_A\|\bm D_Z\|_2^2.
\end{aligned}
\end{equation}
Normalized covariance traces sum to at most one.
Since $0\le c_k\le\bm h_k^H\bm Q_k\bm h_k\le P\|\bm h_k\|_2^2$,
scalar slacks lie in $[0,1]$ and sum to at most $K+1$.
Power bounds the information trace, and the anchor objective bounds the
optimal epigraph trace. Thus $T_A$ bounds the optimal normalized trace.
Subscript $c$ denotes conditioned information coordinates.
Scalings are inverted before certification.

Exact block spectra and equivalent sparse contractions replace dense
Kronecker operations, with the bundle updates retained.
The total/past bundle sizes are
$(\StrongPublicBundleSize,\StrongPublicPastSize)$ for two targets and $(24,8)$
for three targets, both with adaptive proximal weight.
Separate three-scene calibration sets are used for each target count,
with $N_t=64$ for two targets. Both sets are independent of the test scenes.
Selection maximizes the certification success count, with median runtime
breaking ties in the two-target calibration.
The serious/null-step parameters are $\beta=0.25$, $\mu=0.5$.
Master problems use MOSEK 11.1.10 with tolerance $10^{-8}$.

Certificates and the limits of 1000 iterations and 120 seconds are checked
at iteration one and every ten iterations thereafter.
A successful certificate check ends the run.
The time budget can therefore be exceeded by at most one check interval.
Model construction, conversion, and checkpoint communication are timed.
MATLAB startup and warm-up are excluded. Paired method order is alternated.

\section{Additional Numerical Results}
\label{app:additional_experiments}\label{sec:supp_checks}
\flushend
Paired methods use the same channels and sampled information operator.
User channels are generated as
\begin{equation}\label{eq:supp_rician_channel}
 \bm h_k=\sqrt{3/4}\,\bm a(\phi_k)
       +(4N_t)^{-1/2}\bm g_k,\qquad
 \bm g_k\sim\mathcal{CN}(\bm0,\bm I).
\end{equation}
The scattering vectors are independent, with no additional path-loss factor
and $\E\|\bm h_k\|_2^2=1$. Default user centers are equally spaced in
$[-45^\circ,45^\circ]$ for $K>1$ and set to $-20^\circ$ for $K=1$.
Independent uniform $\pm1^\circ$ jitter is added. In the overlap setting,
the centers are $(-15^\circ,15^\circ)$.
Communication-feasible references with power strictly below $P$ were
verified for the primary-timing, controlled-expansion, and task-grid instances.

Primary timing starts before instance construction and ends after
certification, before result serialization. Initialization, all restoration
passes, and all solve attempts and certificate checks are included, whether
successful or unsuccessful. Cases run serially with one BLAS and one MOSEK
thread, except in the thread-count comparison.
Unless stated otherwise, mean $95\%$ confidence intervals use Student's
$t_{n-1}$ over $n$ scenes.

\subsection{Expansion and Solver Comparisons}
\label{app:controlled_expansion}
The effect of the expansion rule is examined with a common initialization
and certificate target.
The controlled expansion study uses $K=2$ and $N_t=64,128,256$.
Posterior draws follow the Latin-hypercube sample-size rule in
Section~\ref{sec:exp_setup} of the main text.
Separated target centers are equally spaced in $[-35^\circ,35^\circ]$.
Two-target cases use $6^\circ$ widths. In overlapping two-target scenes,
user and target centers are $(-15^\circ,15^\circ)$.
The $M=3$ widths and gain means are $(1^\circ,6^\circ,12^\circ)$ and
$(1,0.6,0.3)$. For $M=4$, they are $(1^\circ,3^\circ,6^\circ,12^\circ)$
and $(1,0.8,0.5,0.3)$. Gain CV is $0.5$.
Five scenes per regime are used with nested arrays.

\begin{figure}[!ht]
 \centering
 \includegraphics[width=.9\columnwidth]{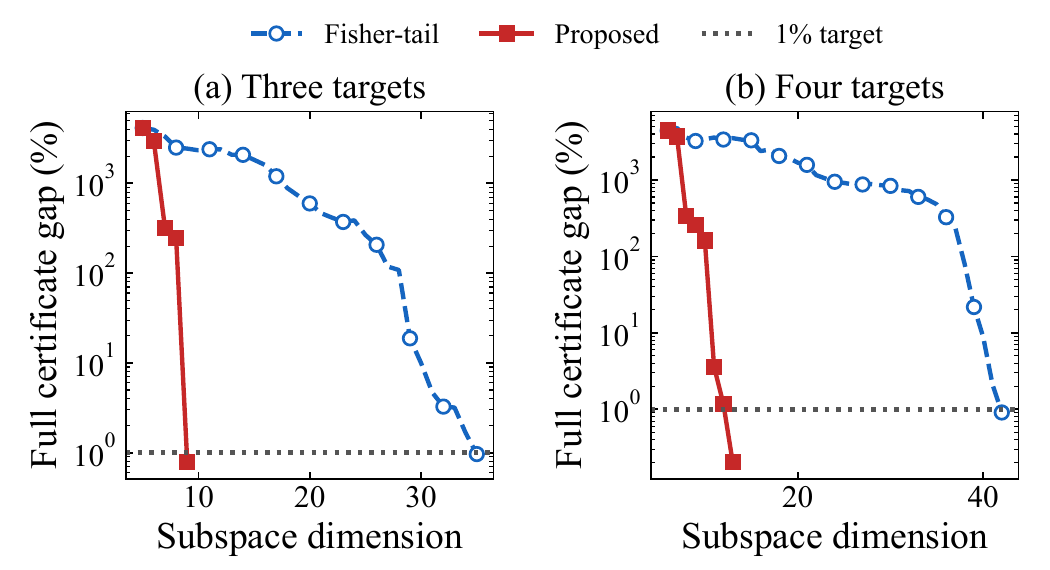}
 \caption{Certificate paths of the proposed method and Fisher-tail expansion.
 Panel (a) repeats Fig.~\ref{fig:exp_expansion_ablation}(b) for comparison with (b).}
 \label{fig:supp_expansion_paths}
\end{figure}

Both rules use the same
timed Fisher ordering and a $1\%$ target. A target of $0.1\%$ is also tested at
$N_t=128$ in the three harder regimes. Tail ratios are computed over jointly
certified pairs. All three-target tail runs at $N_t=256$ and all four-target
tail runs at the $0.1\%$ target exhaust the 120-second budget. Four-target
cases are run up to $N_t=128$. All \StrongStrictSteps{} dynamic expansions satisfy
$f_{{\rm up},+}<L_{\bm U}$, with minimum normalized separation
$\StrongStrictMinSeparation\%$. Fig.~\ref{fig:supp_expansion_paths} shows the
first three- and four-target paths at $N_t=128$.

\begin{figure}[!t]
 \centering
 \includegraphics[width=\columnwidth]{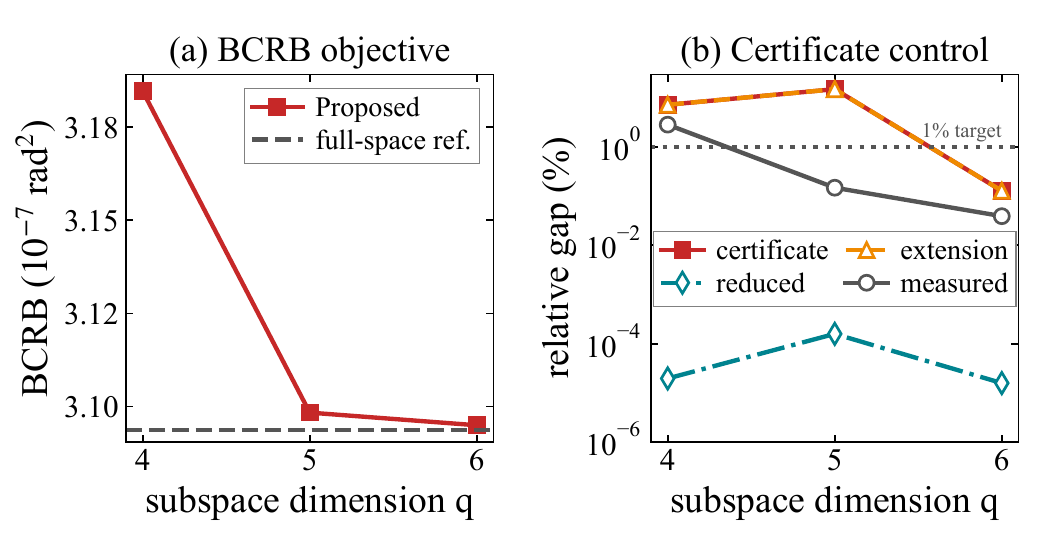}
 \caption{BCRB and certificate evolution during subspace expansion.}
 \label{fig:exp_certificate_trace}
\end{figure}

Fig.~\ref{fig:exp_certificate_trace} follows a predefined angular-overlap
scene at $N_t=64$ with user and target centers at $(-15^\circ,15^\circ)$. The
method visits dimensions four, five, and six. The reduced certificate stays
small throughout, while the extension term identifies when another spatial
direction is needed.

\begin{figure*}[!t]
 \centering
  \begin{minipage}[t]{.29\textwidth}\centering
   \includegraphics[width=\linewidth]{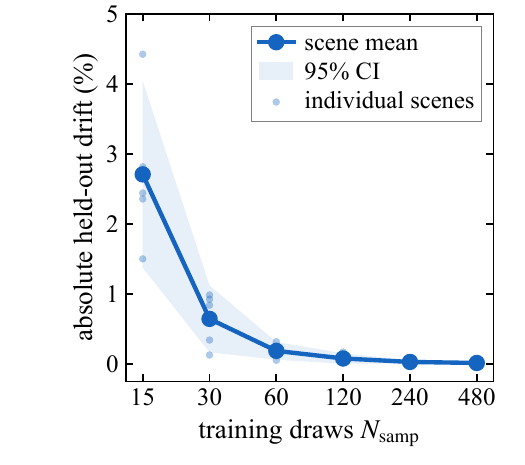}
   \par\footnotesize (a) BCRB drift
  \end{minipage}\hfill
  \begin{minipage}[t]{.29\textwidth}\centering
   \includegraphics[width=\linewidth]{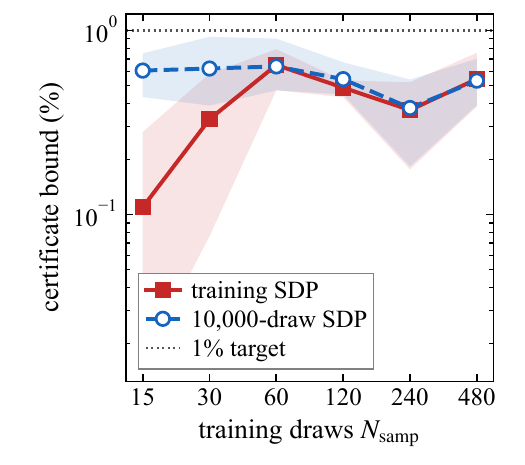}
   \par\footnotesize (b) Sampling certificates
  \end{minipage}\hfill
  \begin{minipage}[t]{.29\textwidth}\centering
   \includegraphics[width=\linewidth]{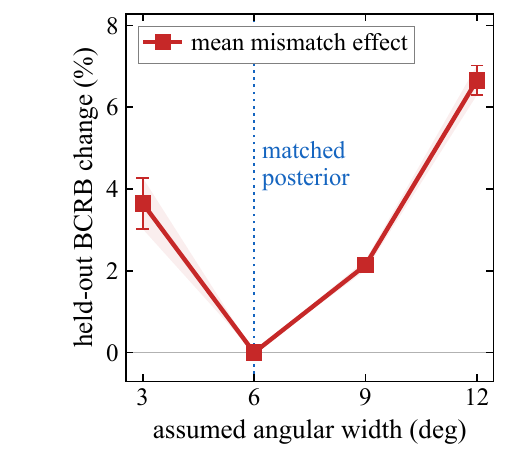}
   \par\footnotesize (c) Width mismatch
  \end{minipage}
 \caption{Posterior sampling (a), (b) and posterior-width mismatch (c).}
 \label{fig:exp_sampling}
\end{figure*}

Table~\ref{tab:exp_baseline_summary} reports results for ten paired $N_t=24$
scenes with widths $(1^\circ,12^\circ)$ and gain means $(1,0.3)$.
Times are medians [$Q_1$, $Q_3$] in seconds over certified runs.
Ratios are medians of scene-wise ratios to Dynamic Fisher.
Fixed-order controls are constructed by appending the initial Fisher,
response-PCA, steering, or DPSS ordering without updating it. Resp-PCA uses
steering vectors and normalized angle derivatives. DPSS~\SuppCite{Slepian1978} sectors are
$\bar\theta_m\pm c\sigma_{\theta,m}$, with $c\in\{1,2,3\}$ calibrated
at $q=6,8$ on three independent calibration scenes. The mean reference gaps
give $c=1$, which is then used for ten separate test scenes.
In these scenes, the proposed method reaches the certificate target with
fewer dimensions and lower median runtime than the fixed-order controls.
\begin{table}[!t]
\centering
\caption{Baseline comparison at a $1\%$ certificate.}
\label{tab:exp_baseline_summary}
{\footnotesize\setlength{\tabcolsep}{2pt}
\begin{tabular}{lcccc}
\toprule
Method & Success & $q_{\rm final}$ & Time [$Q_1$, $Q_3$] & Ratio\\
\midrule
Full-space MOSEK & 10/10 & 24 & 0.109 [0.106,0.117] & 1.42\\
Proposed & 10/10 & 6--7 & 0.076 [0.075,0.085] & 1.00\\
Fisher-tail & 10/10 & 11--13 & 0.219 [0.206,0.240] & 2.80\\
Resp-PCA-tail & 10/10 & 15--17 & 0.406 [0.400,0.458] & 5.34\\
Steering-tail & 10/10 & 11--13 & 0.234 [0.234,0.235] & 3.10\\
Tuned DPSS-tail & 10/10 & 9--10 & 0.151 [0.136,0.164] & 1.83\\
Dynamic Resp-PCA & 10/10 & 6--7 & 0.089 [0.088,0.090] & 1.18\\
Dynamic Steering & 10/10 & 6--7 & 0.088 [0.066,0.090] & 0.93\\
Factorized AL & 10/10 & 6 & 5.381 [3.851,6.587] & 68.71\\
\bottomrule
\end{tabular}}
\end{table}

Table~\ref{tab:supp_initialization} compares Fisher and Resp-PCA initialization
under the same expansion rule, solver policy, and $1\%$ certificate on eight
paired scenes per setting. Mean dimensions and median times are reported.
Fisher initialization uses fewer dimensions in all these settings.
Its runtime benefit is clearer in separated scenes, while the two
initializers have similar runtimes in overlap scenes.
\begin{table}[!t]\centering
\caption{Initialization comparison.}\label{tab:supp_initialization}
{\footnotesize\setlength{\tabcolsep}{3pt}
\begin{tabular*}{\columnwidth}{@{\extracolsep{\fill}}lccccc@{}}\toprule
Scene & $N_t$ & \multicolumn{2}{c}{Fisher} & \multicolumn{2}{c}{Resp-PCA}\\
\cmidrule(lr){3-4}\cmidrule(l){5-6}
 & & $q$ & Time (s) & $q$ & Time (s)\\\midrule
Separated & 64 & 4.88 & 0.132 & 7.00 & 0.177\\
 & 128 & 4.88 & 0.264 & 7.00 & 0.354\\
 & 256 & 4.88 & 1.448 & 7.00 & 1.931\\
 & 512 & 4.88 & 9.549 & 6.88 & 12.775\\
\midrule
Overlap & 64 & 6.00 & 0.168 & 6.88 & 0.172\\
 & 128 & 6.00 & 0.361 & 6.88 & 0.340\\
 & 256 & 6.12 & 1.961 & 7.12 & 1.933\\
 & 512 & 6.12 & 11.944 & 6.75 & 12.652\\
\bottomrule\end{tabular*}}
\end{table}

To assess the choice of probing rank, the AL baseline is run with
$r_s=2,3,7$ and three paired scenes at each of
$N_t=64,128$. All 18 runs meet $1\%$. Rank two gives the lowest median
paired AL/Proposed time ratios, $24.4$--$30.8$, against $34.4$--$65.8$ for rank seven.
The SBMD comparison uses eight paired scenes per setting.
Separated and overlap scenes both use $N_t=64,128,256$.
Three-target scenes use $N_t=128,256$ and the
widths and gain means specified above. Bundle sizes follow the calibration
in Section~\ref{app:public_specbm}.

Performance under random geometry is evaluated with eight scenes per setting.
Centers lie in $[-60^\circ,60^\circ]$, with target separation of at least
$10^\circ$.
All proposed designs are certified.
Table~\ref{tab:supp_nominal_random} compares nominal and certified designs
for two and three targets.
Excesses are normalized by the full-space lower bound. Nominal entries
give median [minimum, maximum].
After shared model construction, median paired speedups are
$\ReceiveRandomSmallRatio$, $\ReceiveRandomMiddleRatio$, and $\ReceiveRandomLargeRatio$
at $N_t=128,256,512$ for $M=2$, and $\ReceiveRandomMultiRatio$ at $N_t=256$ for $M=3$.
At 512 antennas, full-space MOSEK with one, four, and eight threads takes
$\ReceiveThreadsLargeOneRatio$, $\ReceiveThreadsLargeFourRatio$, and
$\ReceiveThreadsLargeEightRatio$ times the single-thread Dynamic Fisher runtime.

\par\smallskip\noindent
\begin{minipage}{\columnwidth}
\centering
\captionof{table}{Random-geometry validation.\label{tab:supp_nominal_random}}
{\footnotesize\setlength{\tabcolsep}{2pt}\renewcommand{\arraystretch}{1.18}
\begin{tabular*}{\columnwidth}{@{\extracolsep{\fill}}lcccc@{}}
\toprule
Scene & $N_t$ & Nominal & \multicolumn{2}{c}{Proposed}\\
\cmidrule(lr){3-3}\cmidrule(l){4-5}
 & & Excess (\%) & $q$ & Max. excess (\%)\\
\midrule
\multirow{3}{*}{Random} & 128 & 17.6 [14.0, 21.6] & 6--8 & 0.42\\
 & 256 & 24.4 [22.3, 26.5] & 5--7 & 0.30\\
 & 512 & 29.0 [27.2, 34.4] & 5--7 & 0.39\\
\midrule
\multirow{2}{*}{Random $M=3$} & 128 & 18.5 [15.8, 21.8] & 9--11 & 0.45\\
 & 256 & 24.5 [22.5, 29.1] & 9--10 & 0.74\\
\bottomrule
\end{tabular*}}
\end{minipage}
\par\medskip

\subsection{Posterior Sampling and Mismatch}
\label{sec:exp_sampling}
Sampling sensitivity is assessed by evaluating fixed trained designs on
independent posterior draws. LHS uses independent random permutations and
within-stratum jitters for each target. Training and evaluation use
separate random streams.
The sample-count rule gives $S=120$ for $M=1$, which is tested only at $N_t=64$
in the task grid.

Independent evaluation reports $(f_E-L_E)/f_E$, using the restored training
design's upper bound $f_E$ and a repaired full-space dual bound $L_E$.
Fresh duals are obtained from an auxiliary SDP on the training basis at
tolerance $10^{-8}$, with refinement at $10^{-10}$ when needed.
The evaluation certificate uses the restored training design and the
repaired auxiliary duals.

Table~\ref{tab:sampling_adequacy} uses five scenes per setting for the separated,
overlap, and three-target cases, with one MC and one LHS design per scene.
Independent evaluation uses 1500 draws (1000 for $M=3$, $N_t=256$).
Sensitivity to the evaluation set is tested by repeating all overlap and
three-target LHS evaluations at the same sample counts.
Larger sets of 3000 draws cover all overlap designs at $N_t=64,128,256$
and four three-target designs at $N_t=128$.
The BCRB change is $100|f_{E,r}(\bm Q)/f_{E_0,0}(\bm Q)-1|$,
where $r$ indexes independent repeats and the denominator is each design's
first evaluation.

The effect of the sampling scheme is examined at equal sample counts in
four pretest scenes, separate from the test sets. Two separated and two
overlap scenes at $N_t=512$ use 480 training and 1500 evaluation draws.
The evaluation certificates are $9.21$--$28.10\%$ for MC and
$0.25$--$1.15\%$ for LHS, supporting the use of stratified sampling.

The sampling study in Fig.~\ref{fig:exp_sampling}(a), (b) uses five paired
$N_t=24$ scenes with $15$--$480$ training draws and 10000 independent
evaluation draws. Panel (a) reports absolute BCRB differences between independent
evaluation and training, normalized by each design's training BCRB.
Panel (b) shows scene means with percentile intervals from 10000 paired
scene-bootstrap resamples.
At 15, 120, and 480 training draws, mean absolute BCRB drifts are
$\RevDriftSmall\%$, $\RevDriftMiddle\%$, and $\RevDriftLarge\%$.
Median independent-sample certificates are $\RevHeldCertSmall\%$,
$\RevHeldCertMiddle\%$, and $\RevHeldCertLarge\%$.

For a population objective $f$ and sampled counterpart $\widehat f$,
suppose uniformly on the feasible set that
$\|\bm J-\widehat{\bm J}\|_2\le\epsilon_J$ and both matrices exceed
$\mu\bm I$, $\mu>0$. The inverse identity gives
$|f-\widehat f|\le M\epsilon_J/\mu^2$.
Let $\Delta_{\rm sampled}$ be an absolute gap certificate satisfying
$\widehat f(\widehat{\bm Q})-\widehat f^\star\le\Delta_{\rm sampled}$. Then
\begin{equation}\label{eq:supp_population_gap}
 f(\widehat{\bm Q})-f^\star
 \le\Delta_{\rm sampled}+2M\epsilon_J/\mu^2.
\end{equation}
For mismatch (Fig.~\ref{fig:exp_sampling}(c)), five scenes use assumed widths $3^\circ,6^\circ,9^\circ,12^\circ$
and evaluation width $6^\circ$. Changes are measured relative to the held-out
BCRB of the design trained at the matched $6^\circ$ width.
The largest mean change is $\ReceiveMismatchMax\%$.

\end{document}